\documentclass[fleqn,usenatbib]{mnras}

\usepackage{newtxtext,newtxmath}

\usepackage[T1]{fontenc}

\DeclareRobustCommand{\VAN}[3]{#2}
\let\VANthebibliography\thebibliography
\def\thebibliography{\DeclareRobustCommand{\VAN}[3]{##3}\VANthebibliography}

\usepackage{graphicx}	
\usepackage{amsmath}	

\usepackage{caption}

\defcitealias{Battaglia22}{B22}
\defcitealias{Pace22}{P22}

\def\eg{{e.g.,\ }}

\usepackage{amsmath}

\def\kms{{\rm\,km \  s^{-1}}}
\def\FeH{{\rm[Fe/H]}}

\def\ione{{~\sc i}}
\def\ii{{~\sc ii}}

\usepackage[dvipsnames,table]{xcolor}

\title[Local Group galaxies seen by DESI DR1]{Confirming membership in Local Group galaxies with the Dark Energy Spectroscopic Instrument Data Release 1}

\author[F. Sestito et al.]{
Federico Sestito$^{1}$\thanks{E-mail: f.sestito@herts.ac.uk} 
\&
 Chiaki Kobayashi$^{1}$
\\
$^{1}$Centre for Astrophysics Research, University of Hertfordshire, Hatfield, AL10 9AB, UK\\
}

\date{Accepted XXX. Received YYY; in original form ZZZ}

\pubyear{\the\year{}}

\begin{document}
\label{firstpage}
\pagerange{\pageref{firstpage}--\pageref{lastpage}}
\maketitle

\begin{abstract}
We use the Dark Energy Spectroscopic Instrument Data Release 1 (DESI DR1)  to identify stellar members of the Local Group dwarf galaxies. We cross-match DESI targets with candidate members that are based on Gaia proper motions, positions, and photometry. The addition of DESI radial velocities enables secure membership determination in 15 systems.
Our results confirm that Gaia-based selection algorithms are effective in minimising foreground contamination.  
Two stars are found to be associated with DES~J0225$+$0304; if this is the case, then it leads to the first determination of the systemic radial velocity (RV$_{\rm{sys}}=-150.0\pm7.0$~km~s$^{-1}$). 
Draco and Sextans are the galaxies with the largest number of members. 
We focus on Sextans and, for the first time with DESI, trace its stellar kinematics to large radii (up to $\sim$10~half-light radii). 
We find that the metal-poor population exhibits a higher velocity dispersion and extends to larger radii, whereas the  metal-rich population is kinematically colder and centrally concentrated. 
The metallicity gradient is steeper in the inner regions of Sextans ($\sim -12\times 10^{-3}$~dex~arcmin$^{-1}$ or $\sim -0.36$~dex~kpc$^{-1}$), while almost no gradient in the outskirts, hinting for an ex-situ halo or for an ``outside-in'' star formation. 
Although DESI [$\alpha$/Fe] ratios for Sextans stars with $\FeH\gtrsim-2.0$ are in line with literature values, those for very metal-poor stars ($\FeH\lesssim-2.0$) present a large scatter and strong anti-correlation with metallicity, warranting a caution for using DESI abundances in this regime. 
With a less strict selection, we identify 8 ultra metal-poor ([Fe/H]~$< -4$) candidates that require higher signal-to-noise ratio spectroscopic observations to determine their metallicities.
\end{abstract}




\section{Introduction}
Nearby dwarf galaxies offer a unique laboratory for understanding galaxy formation and evolution \citep[e.g.][]{Tolstoy09,Bullock17,Simon19}, as only in the Local Group it is possible to derive detailed chemo-dynamical information of {\it individual} member stars. Such low-mass systems could be the building blocks of the Milky Way, or might be analogues of the first galaxies in the early Universe.
Cosmological simulations with the standard $\Lambda-$CDM cosmology seemed to overproduce dwarf satellite galaxies in the Milky Way halo, especially at the low-mass end (the so-called missing satellites problem; e.g., \citealt{Klypin99,Moore99}). This tension has now be solved with high-resolution hydrodynamical simulations including baryonic physics  \citep[e.g.][]{Sales22,SantosSantos26}, as well as with the discovery of new systems thanks to the advent of various surveys, e.g. the Sloan Digital Sky Survey \citep[SDSS,][]{York00,Willman05}, the Dark Energy Survey \citep[DES,][]{DES05}, Pan-STARSS1 \citep[][]{Chambers16}, UNIONS \citep{Smith23,Gwyn25}, the Hyper Suprime-Cam Subaru Strategic Program \citep[HSC-SSP][]{Aihara18}, Gaia \citep{Gaia16,Torrealba19}, and DELVE \citep{DrlicaWagner21}. However, there is another problem related to the density and kinematics of stars in the inner regions of dwarf galaxies, i.e. the core-cusp problem \citep[e.g.][]{Flores94,Navarro96b,Genina18}; the diversity of densities among various dwarf galaxies still remains an open question to be addressed \citep[e.g.][]{Sales22}.

Spectroscopic observations of dwarf galaxies in the Local Group are expensive as these systems are intrinsically faint with low surface brightness, in addition to large distance from us. Hence, the number of observed stars with detailed spectroscopic information is limited and mostly confined to the inner parts of the systems. Thanks to the photometry, the exquisite proper motion, and the positions of stars from the recent data releases of Gaia \citep{Gaia16,GaiaDR2,GaiaDR3}, it is now possible to select candidate members stars with high purity\footnote{Purity is simply defined as the percentage of those stars that are confirmed to be members over the total number of candidates.} \citep[e.g.][]{McVenn2020b,Battaglia22,Pace22,Jensen24, Qi22}, even in the extreme outskirts of galaxies \citep[up to $\sim$12~half-light radii, e.g.][]{Chiti21,Filion21,Yang22,Vitali22,Longeard22,Longeard23,Longeard25,Hayes23,Roederer23,Sestito23scl,Sestito23Umi,Sestito24Sgr,Sestito24Sgr2,Smith23,Waller23,Tau24,Sato25}. 

The relatively small sizes and low masses of dwarf galaxies make them extremely susceptible to not only internal but also external physical processes that determine their morphologies, especially in their outskirts \citep[\eg][and references therein]{Higgs21}. Among the internal processes, those that can influence the morphologies are star formation and  stellar feedback \citep[\eg][]{ElBadry18b}; the external processes include mergers \citep[\eg][]{Deason22}, ram pressure stirring \citep[\eg][]{Kazantzidis11} and stripping \citep[\eg][]{Grebel03}, tidal interaction \citep[\eg][]{Penarrubia08,Fattahi18}, and reionization \citep[\eg][]{Wheeler19}. Therefore, studying the chemo-dynamical properties of individual stars from the inner regions to the extreme outskirts of a dwarf  galaxy is crucial to understand these physical processes that drive galaxy formation and evolution.

To confirm the membership, radial velocities are need to remove foreground and background stars. While Gaia's RV is too shallow, multi-object spectrographs (MOS) with thousands of fibres will drastically improve our knowledge on dwarf galaxies, e.g. WEAVE \citep{WEAVE12,WEAVE24}, 4MOST \citep{4MOST_hrdisk,4MOST_dwarf}, and PFS \citep{PFS,PFSGA}.

The MOS survey we use here is the Dark Energy Spectroscopic Instrument  data release 1 (DESI DR1), which has the primary goal to investigate the nature of the dark energy and the expansion of the Universe \citep[e.g.][]{DESI16,DESIEDR,Poppett24}. The second goal of  DESI, under bright observing conditions, is to observe stars in the Local Group, using its 5,000 fibres \citep{DESIMW,DESIstellarcat,Cooper23}. 

So far radial velocities and metallicities from DESI~DR1 \citep{DESIstellarcat,Cooper23} have been used to study the properties of five stellar structures in the Milky Way halo  \citep{DESIstellarstruc}; to derive the metallicity gradient, the velocity and its dispersion of Draco dwarf galaxy \citep{DESIdraco}; and to investigate the astrophysical J and D factors in Sextans, Draco, and Ursa Minor using stars within $2.5$ half-light radii \cite[in addition to private DESI data,][]{DESIdkm}.

In this paper, we use radial velocities and metallicities from the full dataset of DESI DR1 to confirm or rejects DGs candidate members from \citet{Battaglia22}, and \citet{Pace22}. Section~\ref{sec:data} describes the data used in this work; Section~\ref{sec:crit} reports the selection criteria adopted to confirm membership of DESI targets; Section~\ref{sec:gal} describes the galaxies with at least one star observed by DESI DR1. Section~\ref{sec:gradients} focuses on the Sextans dwarf galaxy, as it is the second-most galaxy with the highest number of members observed by DESI DR1, and for which we are identifying new members in its extreme outskirts (up to $\sim$10~half-light radii). The chemo-dynamical properties of Sextans using DESI DR1 data are discussed, also with respect the literature, and compared to novel galactic chemical evolution models and cosmological hydrodynamical zoom-in simulations. Section~\ref{sec:ump} focuses on the most metal-poor star candidates related to the systems discussed in this paper; lastly, Section~\ref{sec:conclusions} summarises our work.

\section{Data}\label{sec:data}
In this work, we use metallicities and radial velocities from the DESI DR1  stellar catalogue\footnote{The mwsall-pix-iron.fits table is used in this work.} \citep{DESIstellarcat}, from which we use the DESI quantities  extracted with the SP pipeline \citep{Cooper23}, to confirm stellar members in Local Group galaxies. The pipeline uses a code based on \textsc{FERRE}\footnote{The SP pipeline is preferred over the RVS, as we tested values from FERRE code against those from high-resolution spectroscopic (e.g. radial velocity, [Fe/H], surface gravity, and effective temperature) in various works \citep{Arentsen21,Arentsen24,Aguado19,Sestito23,Sestito24Sgr,Sestito24Sgr2}, finding good agreement.} \citep{Allende06}. 
DESI DR1 metallicities \FeH{} have been corrected by the offset described in \citet[][see their Table~4 and Equation~2]{DESIstellarcat}, using a second-order polynomial relation between the uncalibrated \FeH{} and the derived effective temperature of the star\footnote{This correction is not applied to hot stars, i.e. all  stars in M33 and IC1613.}. Uncertainties on the radial velocities have been increased, adding a $0.9\kms$  in quadrature \citep{DESIstellarcat}. 

The DESI DR1  pipeline \citep{DESIstellarstruc} also extracts elemental abundances for carbon, for some $\alpha$-elements (Ca, Mg, Si, Na, Al, and a general $\alpha$), and for Fe-peak elements  (Fe, Cr, Ni, and Ti). However, \citet{DESIstellarcat} concluded that only the elemental abundances of Mg, Ca, and Fe are reliable. In addition, DESI DR1 elemental abundances show a strong anticorrelation with metallicities, especially at the low-metallicity ends \citep[see Figure~6 of][]{DESIstellarcat}.
In this work, we explore DESI DR1 elemental abundances of $\alpha$-elements ([$\alpha$, Mg, Ca/Fe]) for the Sextans dwarf galaxy only, and we compare them with high-resolution spectroscopic values from the literature and galactic chemical evolution models.

\section{Selection criteria of the dwarf galaxy members}\label{sec:crit}
First of all, cross-matched with Gaia DR3, DESI DR1 data are filtered using the following criteria (selection 1): 
\begin{itemize}
    \item Gaia's parallax $\varpi<0.1$ mas in order to remove the majority of Milky Way stars.
    \item Position within 10 half light-radii ($r_h$) from the centre of a given system,  not to cut out any potential members in the extreme outskirts of systems \citep[e.g.][]{Battaglia22,Pace22,Sestito23scl,Sestito23Umi,Jensen24}; values of the systems' $r_h$ are taken from \citet{Battaglia22}.
    \item DESI DR1's surface gravity log($g$)\,$<$\,4 and \textsc{BESTGRID}~!$=$~\textsc{s\_rdesi1} as a fit-quality cut discussed in \citet{DESIstellarcat}. No further cut based on RR\_SPECTYPE or  RVS\_WARN flags is used as they are not included in the SP pipeline.
    \item To avoid background galaxies as discussed in \citet{Riello21}, imposing $\textsc{gaia\_astrometric\_excess\_noise}$\,$<$\,1 and $\textsc{gaia\_phot\_bp\_rp\_excess\_factor} <1.3 + 0.06 \times (\textsc{BP}-\textsc{RP})^2$. 
    \item Radial velocity  uncertainties $\le15$~km~s$^{-1}$.
\end{itemize}

The second set of criteria restricts to stars if they align  with an isochrone in the colour magnitude diagram (CMD), and if they move, in proper motion and line of sight velocity, similarly as the system (selection 2). More specifically, we select stars:
\begin{itemize}
    \item within 0.2 mag in colour (BP $-$ RP), which is well beyond the uncertainties on the Gaia colour, and $2\sigma$ from the distance modulus error in G mag. A very metal-poor  ($\FeH=-2.0$) PARSEC isochrone \citep{Bressan12} with stellar age of 10 Gyr is used in this work;
    \item proper motion within 1.5 mas yr$^{-1}$ of the systemic motion  \citep[values from][]{Battaglia22}, i.e. $\rm{|ppm - ppm_{system}| - \sigma_{ppm} \le 1.5\ mas\ yr^{-1}  }$; and
    \item line-of-sight velocities around the mean systemic velocity \citep[values from][]{Battaglia22, McVenn2020b} and within 5 times of its velocity dispersion, i.e. $\rm{|RV - RV_{system}|  <= 5\times \sigma_{RV, sys} + \sigma_{RV} }$.
\end{itemize}

Finally (selection 3), stars that passed the above two sets of selection criteria are cross-matched with the candidate member catalogues from \citet[][hereafter B22]{Battaglia22} and \citet[][hereafter P22]{Pace22},  imposing a probability to be a member $>50$ percent from their Bayesian algorithms. The algorithms from  \citetalias{Battaglia22} or \citetalias{Pace22} assign a probability to be a member of a given system based on the position on the sky, CMD, and proper motion\footnote{In addition to those quantities and only for some systems, \citetalias{Battaglia22} also uses spectroscopic information. However, differences in membership estimation are minor and not relevant for the scope of this work. For \citetalias{Pace22}, the  \textsc{Mem\_fixed} probability is chosen.}. 
Stars from our selection have a mean uncertainty on the DESI radial velocities of $\sim3.3$~km~s$^{-1}$, while the mean uncertainty on the metallicity is $\sim0.4$ dex.

\section{Galaxies in DESI DR1}\label{sec:gal}
We identified 15 systems from the cross-matching \citetalias{Battaglia22} and \citetalias{Pace22} with DESI DR1 data, including  ultra-faint dwarfs (e.g. Hercules, Coma Berenice), classical dwarf galaxies (e.g. Sextans, Draco), a spiral galaxy (M33) and an irregular dwarf galaxy (IC1613).
Each row of panels in Figure~\ref{Fig:selexample} shows the projected position on the sky in the direction of a given galaxy for which we identified at least one member (first panel), the Gaia colour magnitude diagram (second), the proper motion space (third), and the metallicity vs radial velocity for DESI DR1 stars (fourth).

In the figure, blue small crosses are stars that passed only our selection 1 and thus should be considered as contaminant stars, as their CMD positions, their proper motions, and/or their radial velocities likely offset from those of the system\footnote{Those contaminant stars close to the CMD have RVs or proper motions that differ consistently  from the systemic values, given the selection criteria.}. Black filled circles are those stars that pass our selections 1$+$2 but are not present in either \citetalias{Battaglia22} or \citetalias{Pace22}, and thus new members.
Red and blue circles are those stars passing selection 1$+$2$+$3 from \citetalias{Battaglia22} and \citetalias{Pace22} catalogue, respectively, and are confirmed members. Interestingly, when present, new members (black circles) are located more outwardly than those members from \citetalias{Battaglia22} and \citetalias{Pace22}. This can be explained by the fact that Gaia-only based algorithms might have missed some members in such outskirts of galaxies. These outskirts members can be recovered once spectroscopic information (RV at least) are used as in this paper.

In addition, red  and blue rhombus are stars that  were likely candidate members  (probability $>50$ percent) according to \citetalias{Battaglia22} or \citetalias{Pace22}, respectively, but their DESI DR1 velocity is not compatible with the system's one within 5 times its velocity dispersion. Moreover, known member stars from the Stellar Abundances for Galactic Archaeology Database  \citep[SAGA,][]{Suda08} have been added as a reference in the first and fourth panels (dark green squares). Table~\ref{tab:members} reports the number of new members, those confirmed or rejected candidates from \citetalias{Battaglia22} and \citetalias{Pace22}, as well as the stellar masses of the galaxies studied in this work. 
In the following subsections, we report the candidate members for each of the systems observed in DESI DR1.
The full list of these targets and their IDs will be provided as online material only.

So far, there are only two works that used DESI DR1 in addition to other data to investigate dwarf galaxies in the Local Group. \citet{DESIdraco} focused on Draco only, while \citet{DESIdkm} investigated the dark matter properties in Draco, Sextans and Ursa Minor.  While these works mostly use candidate members from \citetalias{Pace22}, we also use the larger list from \citetalias{Battaglia22}. 
Note that while \citet{DESIdkm} reports the finding of 1003 members in Ursa Minor, adopting the list of \citetalias{Pace22}, we find none. 
This discrepancy is caused by the fact that \citet{DESIdkm} used DESI tertiary data that are not part of the DR1 or of the science verification, and are not currently available.

\begin{figure*}
\includegraphics[width=1\textwidth]{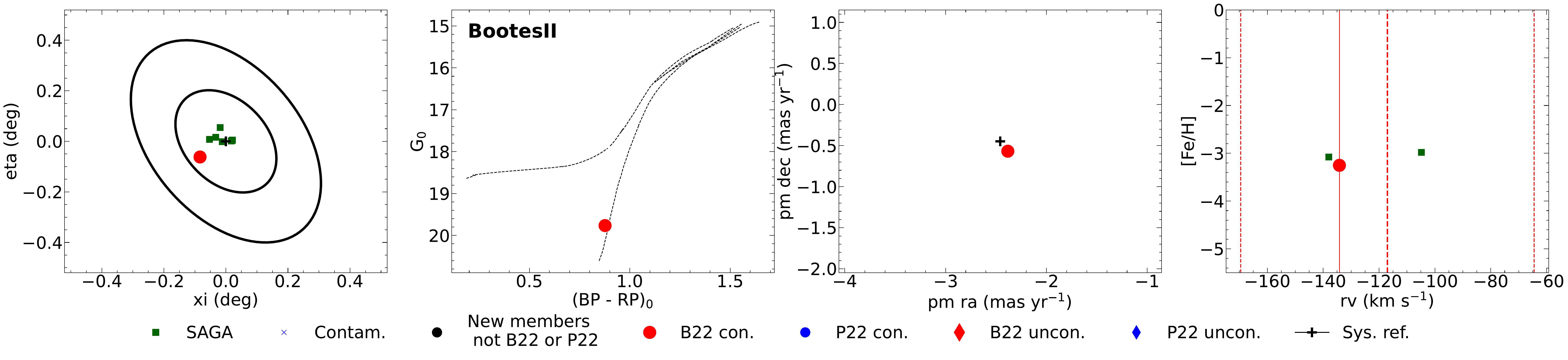}\\
\includegraphics[width=1\textwidth]{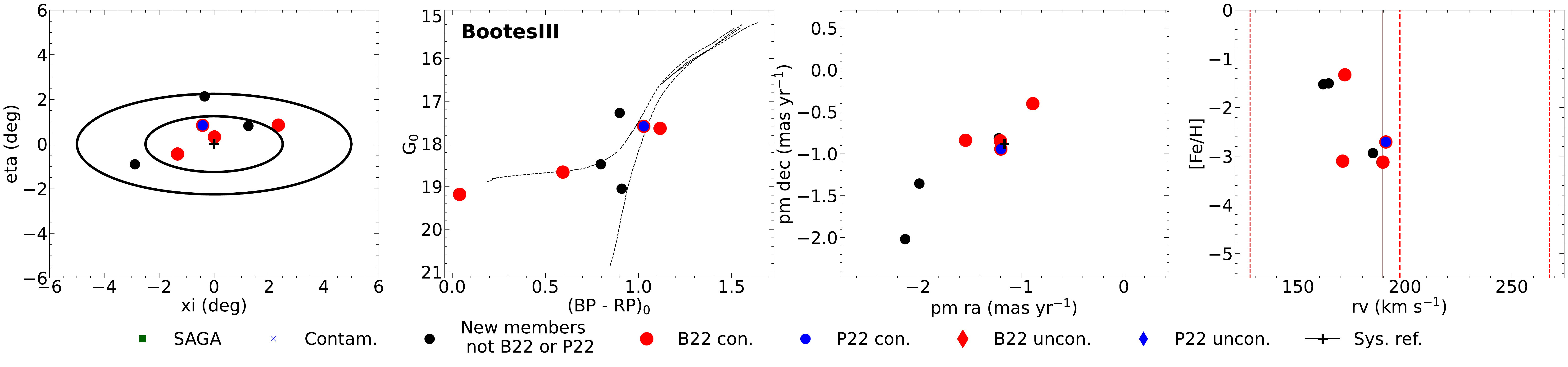}\\
\includegraphics[width=1\textwidth]{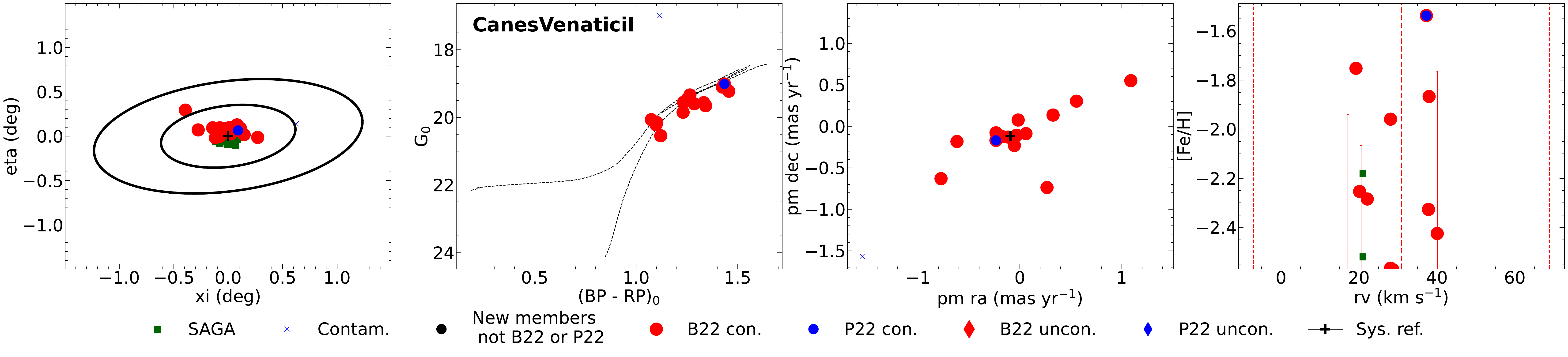}\\
\includegraphics[width=1\textwidth]{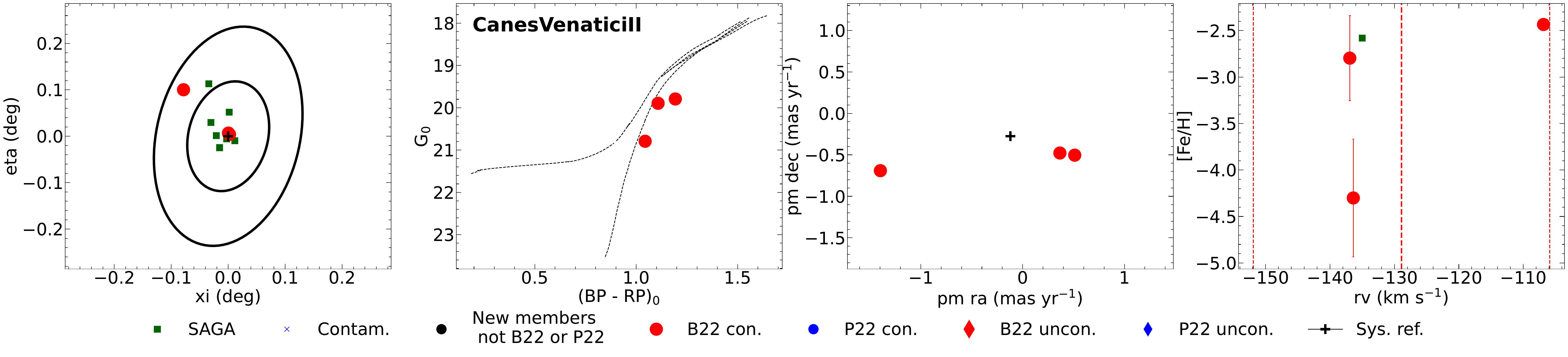}\\
\includegraphics[width=1\textwidth]{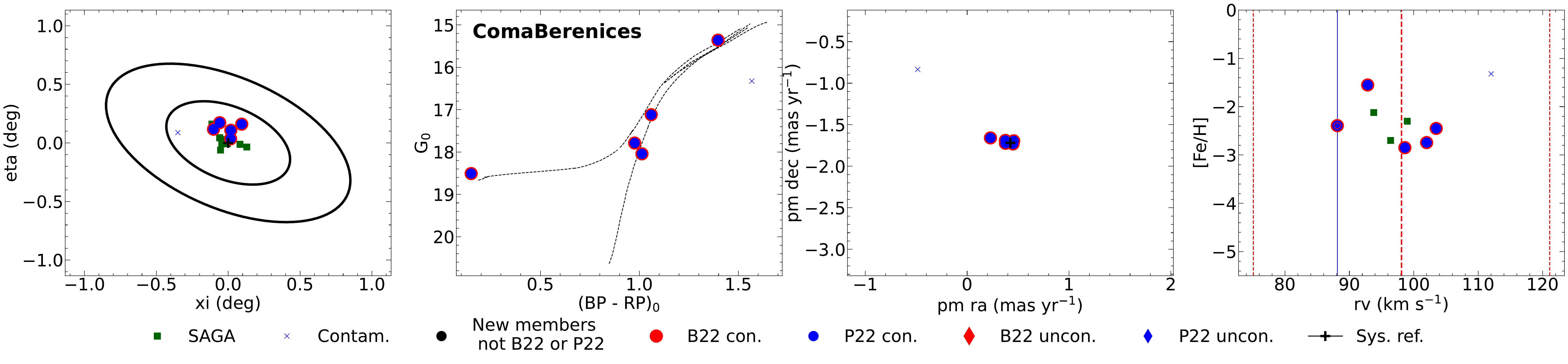}
\caption{Properties of the Local Group galaxies seen in DESI DR1. Fist column: projected on sky position. Two black ellipses represent 5 and 10 half-light radii. Black plus markers denote the position of the system's centre. Second column: Gaia's colour magnitude diagram. An old (10 Gyr) very metal-poor  ($\FeH=-2.0$) PARSEC isochrone \citep{Bressan12} is represented with a black dotted line. Third column: Gaia proper motion space.  Black plus markers denote the systemic proper motion. Forth column: metallicity vs radial velocity space. The three vertical dashed-lines denote the systemic line-of-sight velocity $\pm$5 times its velocity dispersion.}
\label{Fig:selexample}
\end{figure*}

\begin{figure*}\ContinuedFloat
\includegraphics[width=1\textwidth]{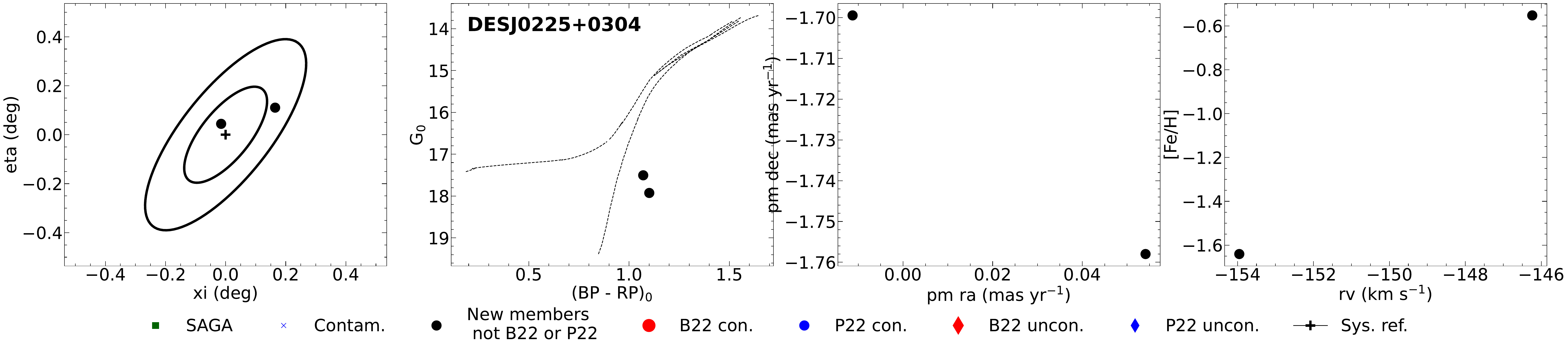}\\
\includegraphics[width=1\textwidth]{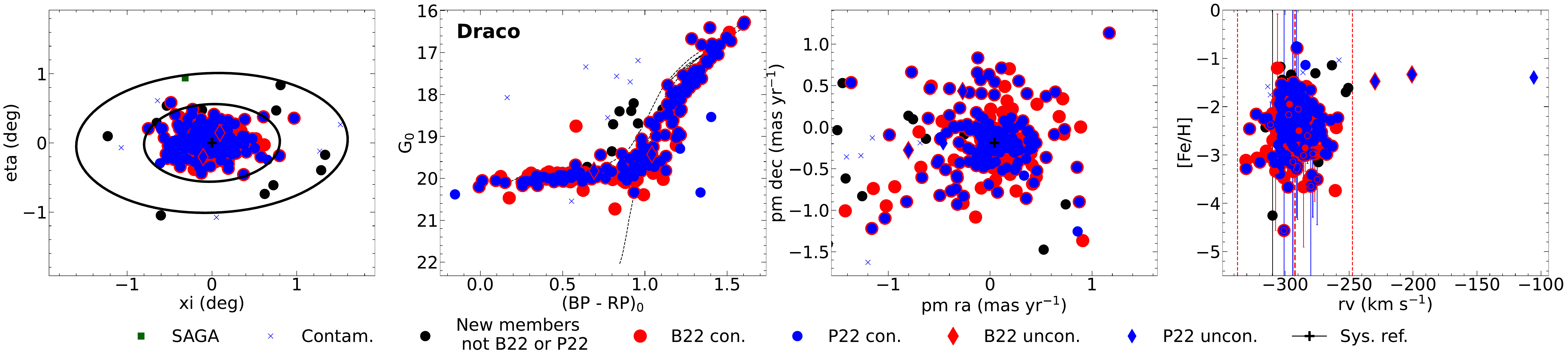}\\
\includegraphics[width=1\textwidth]{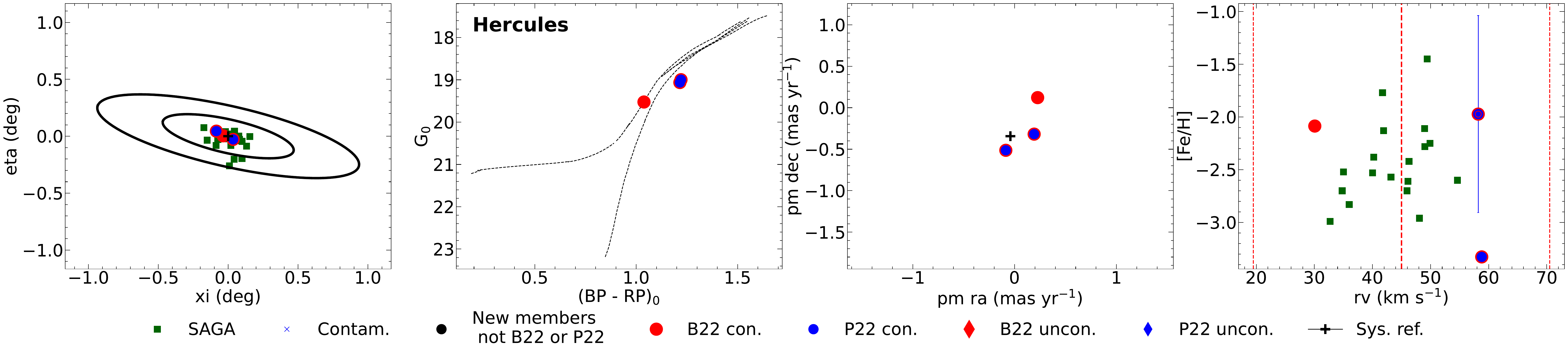}\\
\includegraphics[width=1\textwidth]{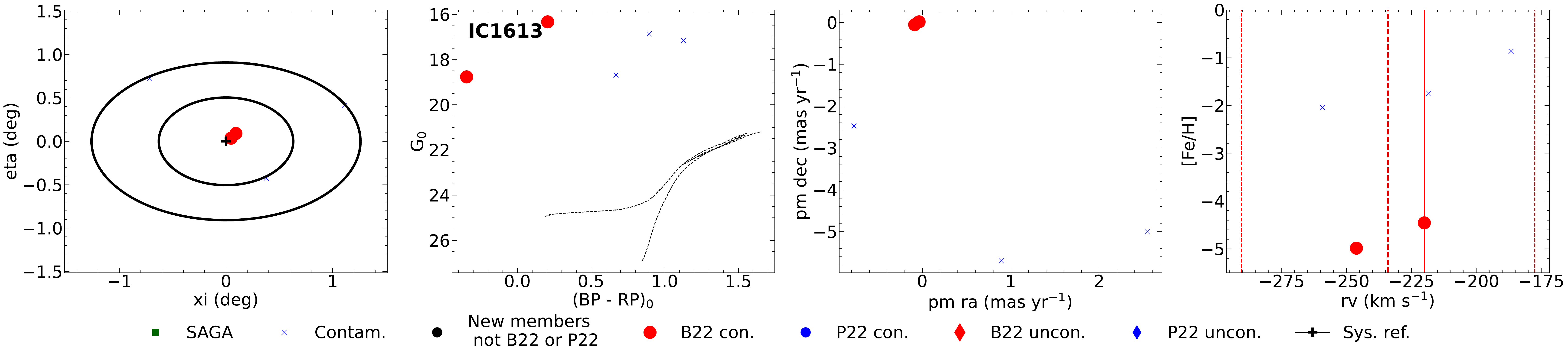}\\
\includegraphics[width=1\textwidth]{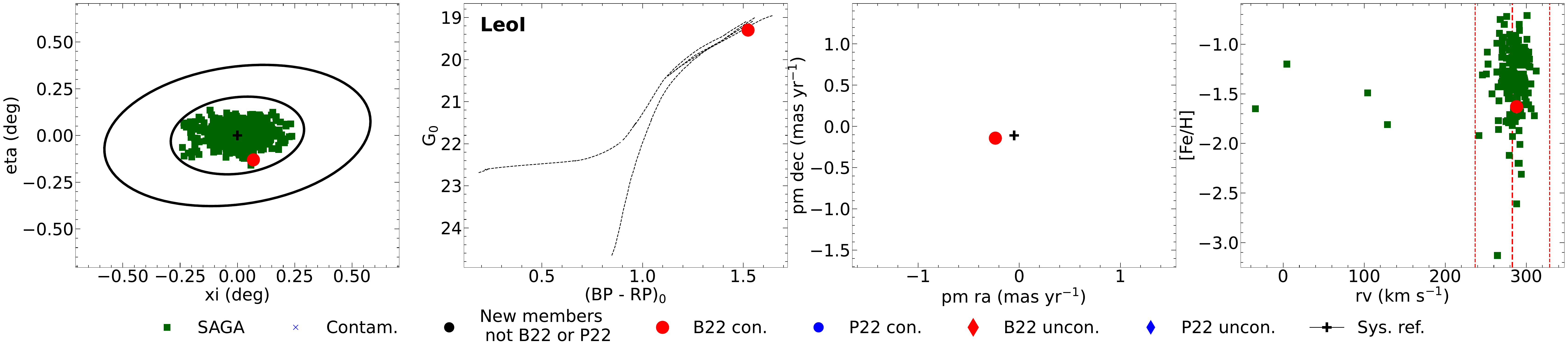}
\caption{(continued) Properties of the Local Group galaxies seen in DESI DR1.}
\label{Fig:selexample}
\end{figure*}

\begin{figure*}\ContinuedFloat
\includegraphics[width=1\textwidth]{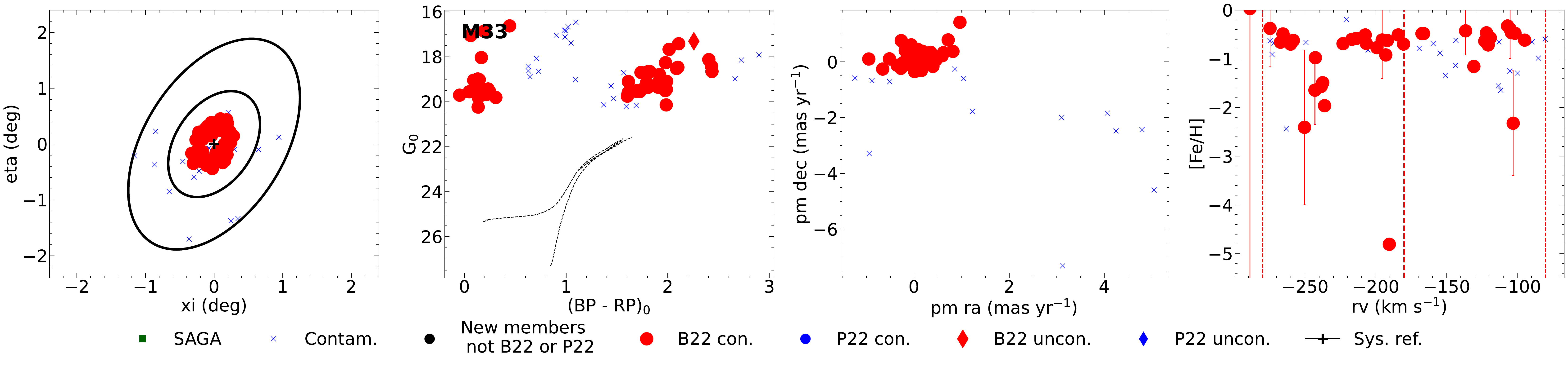}\\
\includegraphics[width=1\textwidth]{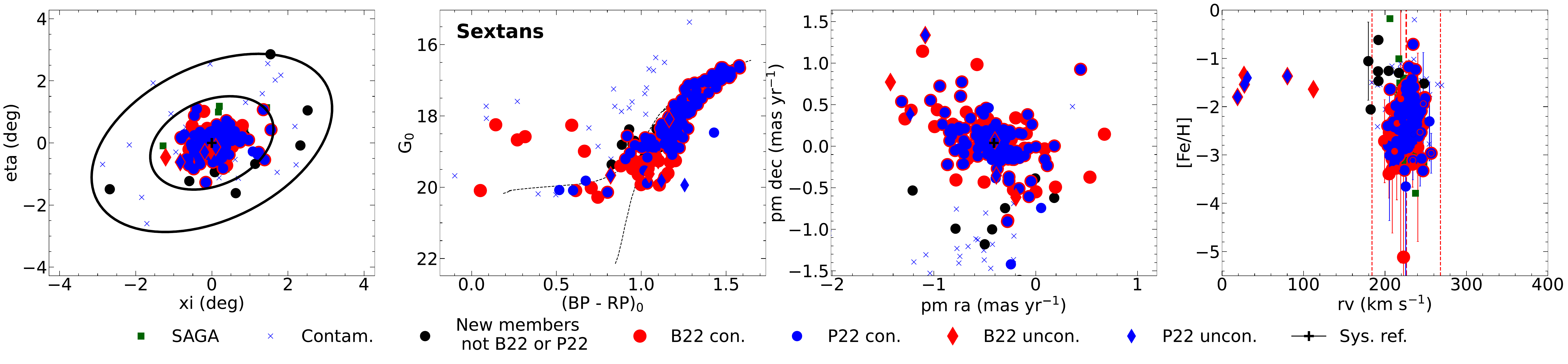}\\
\includegraphics[width=1\textwidth]{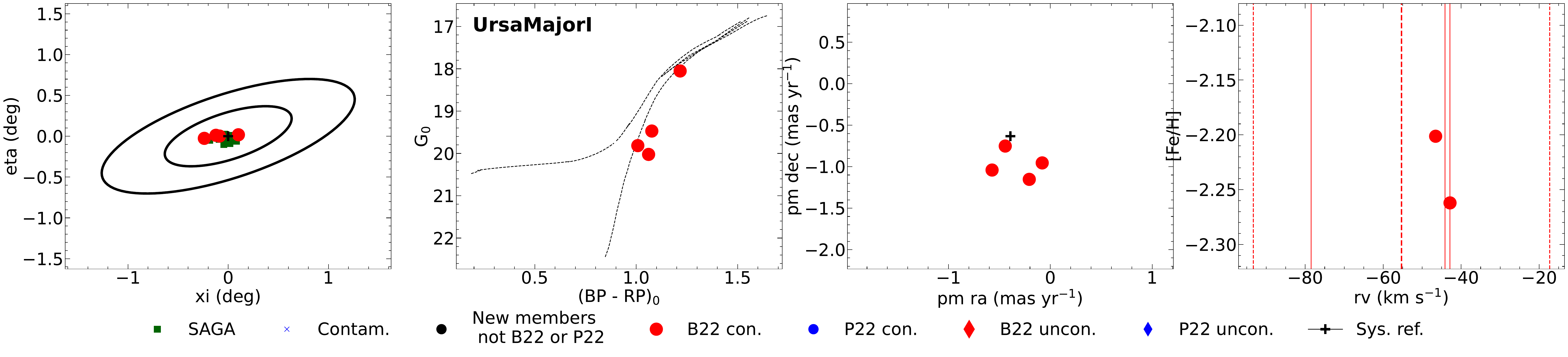}\\
\includegraphics[width=1\textwidth]{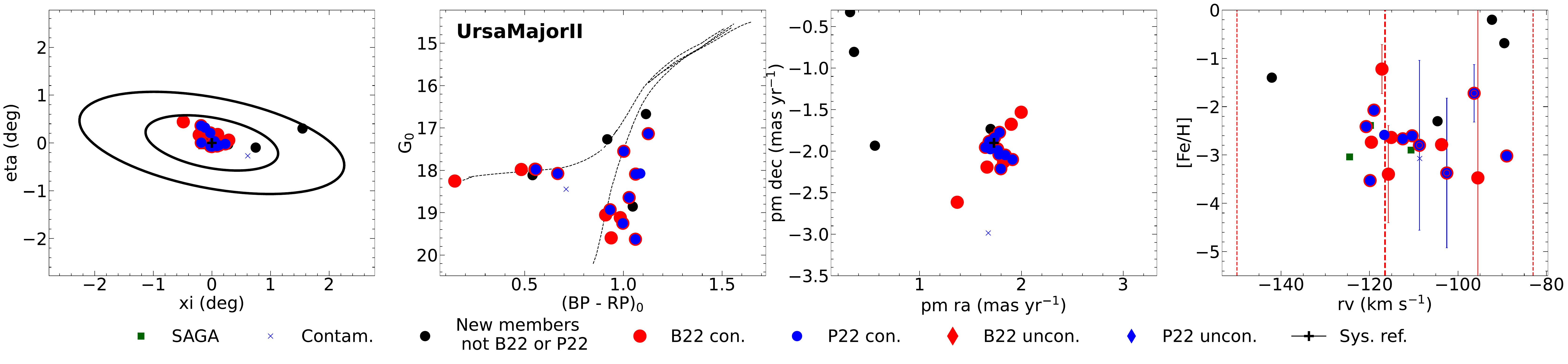}\\
\includegraphics[width=1\textwidth]{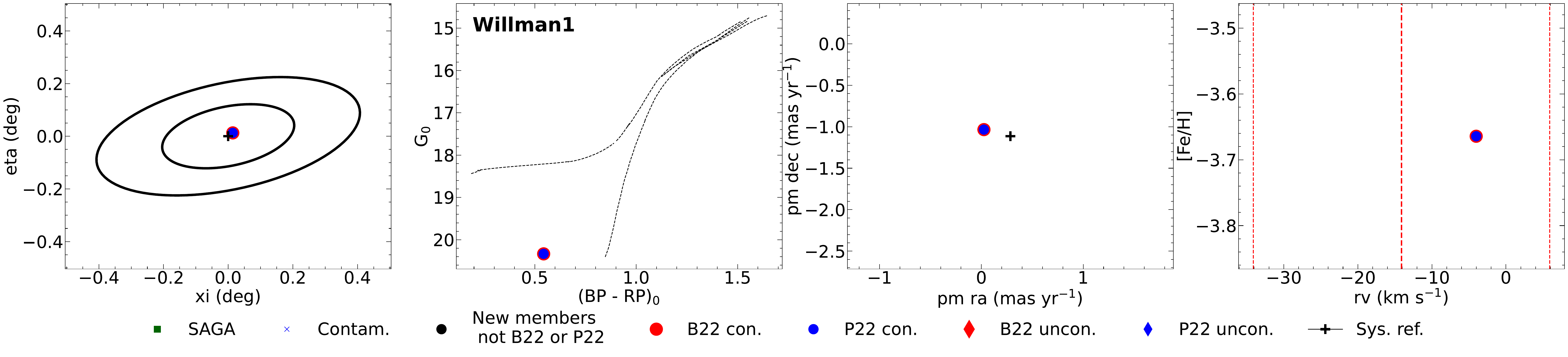}
\caption{(continued) Properties of the Local Group galaxies seen in DESI DR1.}
\label{Fig:selexample}
\end{figure*}

\begin{table*}
    \centering
    \begin{tabular}{ccccccccc}
         Galaxy & M$_{\star}$& New members  & Confirmed  & Confirmed  & Confirmed & Unconfirmed & Unconfirmed & Unconfirmed \\
          & ($10^4$ M$_{\odot}$) & not in B22 nor P22 & B22 & P22 & B22 and P22 & B22 & P22  & B22 and P22\\ \hline
BootesII & 0.1 & 0 & 1 & 0 & 0 & 0 & 0 & 0 \\ \hline
BootesIII & tidally disrupted & 3 & 4 & 1 & 1 & 0 & 0 & 0 \\ \hline
CanesVenaticiI & 23 & 0 & 14 & 1 & 1 & 0 & 0 & 0 \\ \hline
CanesVenaticiII & 0.79 & 0 & 3 & 0 & 0 & 0 & 0 & 0 \\ \hline
ComaBerenices & 0.37 & 0 & 5 & 5 & 5 & 0 & 0 & 0 \\ \hline
DES~J0225$+$0304 & N/A & 2 & 0 & 0 & 0 &0 &0 &0 \\ \hline
Draco & 29 & 12 & 212 & 161 & 157 & 2 & 3 & 2 \\ \hline
Hercules & 3.7 & 0 & 3 & 2 & 2 & 0 & 0 & 0 \\ \hline
IC1613 & 10,000 & 0 & 2 & 0 & 0 & 0 & 0 & 0 \\ \hline
LeoI & 550  & 0 & 1 & 0 & 0 & 0 & 0 & 0 \\ \hline
M33 & 300,000 & 0 & 49 & 0 & 0 & 1 & 0 & 0 \\ \hline
Sextans & 44 & 9 & 144 & 117 & 110 & 5 & 4 & 3 \\ \hline
UrsaMajorI & 1.4 & 0 & 4 & 0 & 0 & 0 & 0 & 0 \\ \hline
UrsaMajorII & 0.41 & 4 & 15 & 10 & 9 & 0 & 0 & 0 \\ \hline
Willman1 & 0.1 & 0 & 1 & 1 & 1 & 0 & 0 & 0 \\ \hline
    \end{tabular}
    \caption{Number of confirmed and rejected members for each identified galaxy, cross-matching DESI DR1 with the candidates in \citetalias{Battaglia22} and \citetalias{Pace22}. Stellar masses are from \citet[][and references therein]{Mcconnachie12}, with the exception of M33 \citep{Vandermarel12}. BootesIII is strongly  influenced by tidal forces and no stellar mass value is available \citep{Carlin18}.}
    \label{tab:members}
\end{table*}

\subsection{Bootes II}
There is only one candidate member form \citetalias{Battaglia22} that has been observed by DESI DR1 and we confirm its membership given its radial velocity (RV). Recently, \citet{Longeard25} confirmed the membership of 9 candidates in Bootes~II, using FLAMES spectrograph, 8 of which had a high membership probability in \citetalias{Battaglia22}. However, only one star is in DESI DR1, which is the red circle in Figure~\ref{Fig:selexample}.

\subsection{Bootes III}
There are 4 confirmed members from \citetalias{Battaglia22}, one of which is also in \citetalias{Pace22}.  In addition, there are 3 stars that are not likely members in either \citetalias{Battaglia22} or \citetalias{Pace22}, but are found to be confirmed members, given their DESI DR1 RVs.

\subsection{Canes Venatici I}
There are 14 stars from \citetalias{Battaglia22} confirmed to be members of Canes Venatici I, including only one from \citetalias{Pace22}; the rest of the candidates in \citetalias{Pace22} have a very low probability to be members.

\subsection{Canes Venatici II}
Three likely candidate members from \citetalias{Battaglia22} are confirmed to be members with DESI DR1 RVs.

\subsection{Coma Berenice}
All the 5 likely candidate members, that are in both \citetalias{Battaglia22} and  \citetalias{Pace22}, are confirmed to be members.

\subsection{DES~J0225$+$0304}
DES~J0225$+$0304 is a system, which is associated with the Sagittarius stellar stream, and, so far, there is no estimate for its systemic radial velocity \citep[e.g.][]{Luque17}. Recently, \citet{McVenn2020b} derived the systemic proper motion using Gaia EDR3.

We find two stars, in the direction of DES~J0225$+$0304, observed by DESI DR1, which are not in the catalogues of \citetalias{Battaglia22} nor  \citetalias{Pace22}. These stars have similar proper motions and DESI DR1 radial velocities. Their proper motion values are within $\sim\!1.8\sigma$ and $\sim\!1.5\sigma$ from the systemic values in right ascension and declination reported by \citet{McVenn2020b}, respectively. If these stars are truly members of DES~J0225$+$0304, then we obtain, for the first time, a systemic radial velocity of RV$_{\rm{sys}}=-150.0\pm7.0$~km~s$^{-1}$.

\subsection{Draco}
We confirm 212 and 161 stars from \citetalias{Battaglia22} and \citetalias{Pace22}, respectively, with 157 in common, to be members of Draco. On the other hand, we reject 2 and 3 stars from the two catalogues (\citetalias{Battaglia22} and  \citetalias{Pace22}, respectively), out of which 2 candidates are in common. Given the DESI DR1 RVs, we add 12 more member stars to Draco.

Recently, the DESI DR1 view of Draco was also discussed in \citet{DESIdraco}.
Their work identified 155 members using astrometry, radial velocities and metallicities. They found 8 stars located outside the tidal radius (Stars 1-8 in their Table~3). With our membership selection (see the criteria $1+2+3$ in Section~\ref{sec:crit}), either using \citetalias{Battaglia22} or \citetalias{Pace22}, we confirm the membership of 5 out of the 8 extra-tidal stars.
Moreover, applying our selection criteria $1+2$, we can recover 11 stars that were neither in \citetalias{Battaglia22} nor \citetalias{Pace22}. Only one of these was found to be an extra-tidal star in \citet{DESIdraco}. Therefore, we can identify more extra-tidal stars combining all the selection criteria, including 6 out of the 8 extra-tidal stars in \citet{DESIdraco}.

\citet{DESIdraco} also derived the Draco's mean line of sight velocity, the velocity dispersion and the mean metallicity of the system, finding values compatible with previous attempts \citep{Kirby11,Kirby13,Munoz18}. In addition, the metallicity gradient is steeper within the first half-light radius, while it flattens in the outskirts of the system \citep{DESIdraco}.

\subsection{Hercules}
Three candidate members from \citetalias{Battaglia22} are confirmed to be members. Two of them are also present in \citetalias{Pace22}. \citet{Longeard23} identified 3 new members using the AAOmega spectrograph. None of these stars are present in  \citetalias{Battaglia22}, \citetalias{Pace22}, nor in DESI DR1.

\subsection{IC1613}
IC1613 is an irregular dwarf galaxy, that includes RR Lyrae variable stars and Wolf-Rayet stars, and thus it is not possible to fit its CMD with an old isochrone, as discussed bs \citetalias{Battaglia22}. Therefore, they did not apply the CMD selection, neither do we. We confirm the membership of the two \citetalias{Battaglia22} candidate members, given their RVs.

\subsection{Leo I}
The only \citetalias{Battaglia22} candidate member is confirmed to be a member, given its DESI DR1 RV.

\subsection{M33}
Similarly to IC1613, M33 hosts young stars and, as discussed in \citetalias{Battaglia22}, the CMD cannot be used to select candidate members. We confirm 49 members from \citetalias{Battaglia22}, while  stars are rejected given their RVs.

\subsection{Sextans}
We confirm 144 and 117 stars from \citetalias{Battaglia22} and  \citetalias{Pace22}, respectively, with 110 in common, to be members of Sextans. On the other hand, we reject 5 and 4 stars from the two catalogues (\citetalias{Battaglia22} and  \citetalias{Pace22}, respectively), out of which 3 rejected candidates are in common. Given the DESI DR1 RVs, we add 9 more member stars to Sextans. DESI DR1 data have also been recently used by \citet{DESIdkm} to investigate the astrophysical J and D factors in Sextans. They found 240 Sextans members starting from the \citetalias{Pace22} candidates list. This number is double than our selection from \citetalias{Pace22} or from \citetalias{Battaglia22}. However, we note that, while \citet{DESIdkm} remove stars with half-light radius $>2.5~r_h$, we are able to find new members out to  $\sim10~r_h$.

\subsection{Ursa Major I}
4 candidate members from \citetalias{Battaglia22} are confirmed to be members, given their RVs from DESI DR1.

\subsection{Ursa Major II}
9 stars are candidate members from both  \citetalias{Battaglia22} and  \citetalias{Pace22}, all of them are confirmed members. In addition to that, 6 stars from  \citetalias{Battaglia22}  and one from  \citetalias{Pace22} are confirmed to be members. 4 stars do not pass the probability cuts from those catalogues, however, their radial velocities are in agreement with those of Ursa Major II, therefore, likely to be members. 

\subsection{Willman I}
The only candidate member, which is present in both \citetalias{Battaglia22} and  \citetalias{Pace22} catalogues, is confirmed to be a member.

\subsection{Gaia candidate membership success}
There are various Gaia-based algorithms in the literature to provide a probability to be a member of a given system \citep[e.g.][]{McVenn2020b,Battaglia22,Pace22,Qi22,Jensen24}.
Given the chosen membership probability threshold in this work, we find a purity of $\gtrsim96$ percent (see Table~\ref{tab:members}). Therefore, algorithms based on the exquisite Gaia data are able to provide high-purity membership, which can also be increased if stricter cuts on the probability are applied, although reducing the completeness and size of the sample.

We note that, for most of the systems studied in this work, the algorithm from \citetalias{Battaglia22} provides a higher number of candidates than those from \citetalias{Pace22}, when  adopting the same cut on probability. We want to stress out that different algorithms might have a different efficiency to weed out the foreground contamination, as well as a different optimisation for reaching the outskirts of the systems.
Confirmation with ground-based MOS is crucial to explore the halo of dwarf galaxies, especially their outskirts.

\section{Chemo-dynamical properties of Sextans}\label{sec:gradients}
In this section, we focus on the chemo-dynamical properties of Sextans, which is the galaxy with the largest number of new members after Draco, which is already studied in other works \citep{DESIdraco}. DESI radial velocities and metallicities are used to derive the line-of-sight systemic velocity, the velocity dispersion, and the metallicity gradient of Sextans, while DESI elemental abundances for $\alpha-$elements in Sextans are compared with high-resolution spectroscopic data from the literature, our galactic chemical evolution models \citep[e.g.,][]{Kobayashi20b}, and our chemo-dynamical simulations \citep[e.g.,][Sestito et al., in prep.]{Vincenzo20}.

\subsection{Systemic velocities and metallicities}

\begin{figure*}
\includegraphics[width=1\textwidth]{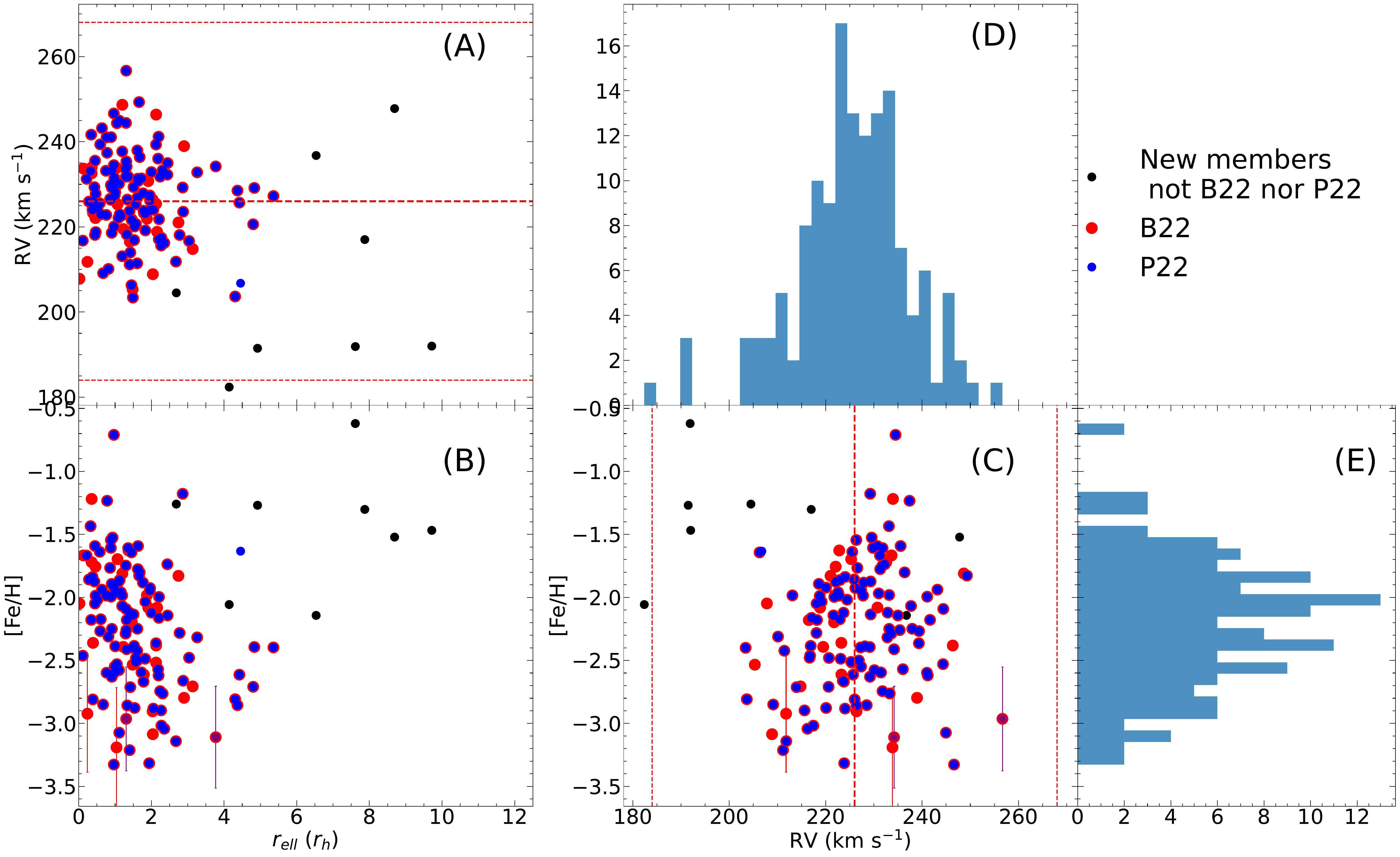}
\caption{Radial velocities and  metallicities in Sextans. Panel A: RVs as a function of the projected elliptical distance. Panel B: \FeH{} as a function of the projected elliptical distance. Panel C: \FeH{} vs RVs. Panel D: RVs' histogram. Panel E: metallicities' histogram. Stars with large uncertainties on the metallicities, $\sigma_{\FeH} > 1.0$~dex, have been removed. Stars with $\sigma_{\FeH} > 0.5$ dex are shown with errorbars.}
\label{Fig:sextans}
\end{figure*}

\begin{table*}
    \centering
    \begin{tabular}{lcccccc}
         Selection & Sample & Sample size &  Systemic RV  & Systemic $\sigma_{RV}$ & Systemic \FeH  & Systemic $\sigma_{\FeH}$ \\
          & & & (km~s$^{-1}$)  & (km~s$^{-1}$) & (dex) & (dex) \\ \hline
        B22 + P22 + New & All & 160 & $ 225.04 ^{+ 0.98 }_{- 0.98 }$ & $ 11.17 ^{+ 0.78 }_{- 0.72 }$  & $ -2.20 ^{+ 0.04 }_{- 0.04 }$ & $ 0.48 ^{+ 0.03 }_{- 0.03 }$\\  \\
         B22 + P22 + New & [Fe/H]~$<-2.0$& 102 & $ 224.82 ^{+ 1.20 }_{- 1.20 }$ & $ 10.79 ^{+ 1.00 }_{- 0.89 }$ & & \\  \\
          B22 + P22 + New &  [Fe/H]~$>-2.0$ & 58 & $ 225.41 ^{+ 1.69 }_{- 1.69 }$ & $ 12.07 ^{+ 1.44 }_{- 1.24 }$ & &  \\  \\ \hline
           B22 + P22 & All  & 151 & $ 226.12 ^{+ 0.82 }_{- 0.84 }$ & $ 8.87 ^{+ 0.71 }_{- 0.63 }$ & $ -2.24 ^{+ 0.04 }_{- 0.04 }$ & $ 0.45 ^{+ 0.03 }_{- 0.03 }$ \\  \\
            B22 + P22 & [Fe/H]~$<-2.0$ & 100 & $ 225.16 ^{+ 1.11 }_{- 1.11 }$ & $ 9.69 ^{+ 0.94 }_{- 0.84 }$ & & \\ \\
             B22 + P22 &  [Fe/H]~$>-2.0$ & 51 & $ 227.71 ^{+ 1.18 }_{- 1.18 }$ & $ 7.45 ^{+ 1.05 }_{- 0.90 }$ & &  \\ \hline \hline
             \citet{DESIstellarcat} &  &   &  &  &  $-2.24^{+0.03}_{-0.04}$ & $0.44^{+0.03}_{-0.03}$ \\ \\
             \citet{Battaglia22} &  &   &   $226.0\pm0.6$ &  $8.4\pm0.4$ &  $-1.90 \pm 0.01$ & $0.6$  \\   \hline
    \end{tabular}
    \caption{Systemic velocities, systemic mean metallicity, and their dispersion for the various samples. Addd values fromm k26 and B11}
    \label{tab:dispersion}
\end{table*}

Panels of Figure~\ref{Fig:sextans} show the radial velocities (RVs) and metallicities ([Fe/H]) in Sextans for those member stars found  in \citetalias{Battaglia22} (red circles), in \citetalias{Pace22} (blue circles), and in DESI alone (black circles), with uncertainty on metallicity $\sigma_{\FeH} < 1.0$ dex.
Panels A and B of Figure~\ref{Fig:sextans} display RVs and [Fe/H] as a function of the projected elliptical distances, normalised by the half light radius \citepalias[$r_{\rm h}=21.4$ arcmin, or $\sim0.7$ kpc,][]{Battaglia22}. Panel C shows RV vs [Fe/H], while panels D and E represent the histograms on these two quantities. In panels B and C, stars with  $\sigma_{\FeH} \geq 0.5$ dex are shown with their errorbars.

Table~\ref{tab:dispersion} reports the systemic line-of-sight velocity, the systemic mean metallicity and their systemic dispersions. These value are derived with a  Markov chain Monte Carlo, based on the Metropolis-Hastings algorithm \citep{Hastings70}, similarly to what adopted by \citet{Longeard20}. As this is embedded in a Bayesian framework, the adopted  prior is simply a step function of the expected radial velocities ($150<\rm{RV}<300$ km s$^{-1}$), of the velocity dispersion ($\sigma_{\rm{RV}}<30$ km s$^{-1}$), of the expected mean metallicity ($-5\leq\FeH\leq-1$), and of the metallicity dispersion ($\sigma_{\rm{\FeH}}<0.5$ dex). The likelihood is a Gaussian distribution, which is centred on the systemic RV (or mean $\FeH$), and with a dispersion equal to the sum in quadrature of the intrinsic dispersion of systemic RV (or mean \FeH) and the uncertainties on the stellar RVs (or \FeH).

Table~\ref{tab:dispersion} lists two sets of values: 
one using  member stars from \citetalias{Battaglia22} plus \citetalias{Pace22} [B22~$+$~P22], and another also including our new members from DESI alone [B22~$+$~P22~$+$~New]. 
For the [B22~$+$~P22] sample, the derived systemic mean metallicity and its uncertainty are identical to those reported by the DESI DR1 stellar catalogue \citep[Table~7 of][]{DESIstellarcat}. The systemic metallicity dispersion  is also in agreement within $\sim0.24\ \sigma$ from the value in  \citet{DESIstellarcat}. Adding our new members, for the  [B22~$+$~P22~$+$~New] sample, our systemic mean metallicity and its dispersion are still in agreement within $0.80\ \sigma$ and $0.94\ \sigma$ from \citet{DESIstellarcat}, respectively.

The systemic RV and its dispersion are also derived  splitting the members according to their metallicities: the very metal-poor (VMP) group with $\FeH<-2.0$ and the  metal-rich (MR) with $\FeH>-2.0$. This value for the threshold has been chosen to arbitrary divide those stars polluted by only Type II supernova from those in which Type Ia supernovae are also contributing \citep[e.g][]{Theler20}.

The systemic velocity and the velocity dispersion obtained in this work, either for all three sources [B22~$+$~P22~$+$~New] or only two sources [B22~$+$~P22], are in agreement with the literature values from \citet{Battaglia11}. We notice that the B22~$+$~P22 values are closer to those from \citet{Battaglia11} than the other sample that includes also black circles (New).

Similar systemic radial velocities, i.e. within $1\sigma$, are also obtained for the VMP and the MR stars, separately. For the velocity dispersion, we note and interesting trend; the MR $\sigma_{\rm{RV}}$ for [B22~$+$~P22~$+$~New] is larger than the VMP value, although we find the opposite behaviour for [B22~$+$~P22]. As a result, the two VMP values from the different datasets are compatible within $1\sigma$, while $\sigma_{\rm{RV}}$ differs only for the MR stars. The discrepancy between the two datasets (B22~$+$~P22~$+$~New vs B22~$+$~P22) for the MR stars might be driven by some relatively metal-rich stars ($-1.5\lesssim\FeH\lesssim-0.5$) that are also located in the outskirts of Sextans, all of them are New (black circles), hence they are not in \citetalias{Battaglia22} nor in \citetalias{Pace22} datasets. Curiously, there are only 8 member stars\footnote{One of the 9 new identified members does not pass the stricter cut on the metallicity uncertainties applied in this Section.} newly identified by DESI (black circles) and almost half of them appear to be related to the dispersion velocity discrepancy.

\subsection{Velocity and metallicity profiles: comparison with observations}

\begin{figure}
\includegraphics[width=0.5\textwidth]{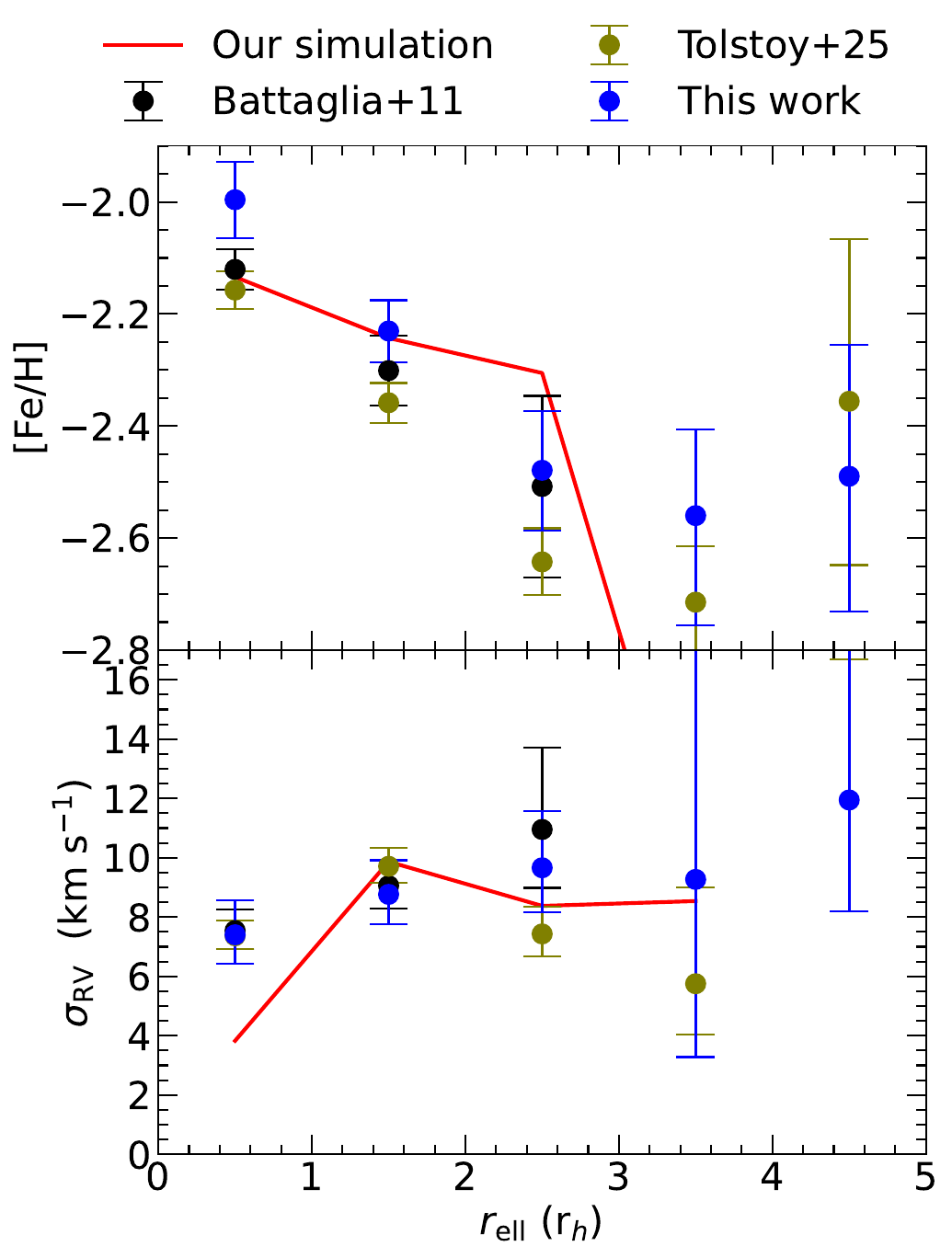}
\caption{Mean metallicity (top) and velocity dispersion (bottom) as a function of the projected elliptical distance in Sextans (half-light radius units). Blue circles mark the median values calculated with the sample from \citetalias{Battaglia22} and \citetalias{Pace22}. 
Black and olive circles represent the median values calculated with the samples from \citet{Battaglia11} and \citet{Tolstoy25}, respectively. Red lines represent the values derived from our simulated dwarf satellite galaxy, with stellar mass of $4.6\times 10^5$~M$_{\odot}$ (similar to Sextans, $\sim4.4\!\times 10^5$~M$_{\odot}$), extracted from a cosmological `zoom-in' hydrodynamical simulation of the Local Group (Sestito et al., in prep.). The mean metallicity of the simulated system has been shifted by $+0.5$ for a better visualisation of the data, while its shape remains intact. No shift has been applied to the velocity dispersion.}
\label{Fig:disp_vel}
\end{figure}

The two panels of Figure~\ref{Fig:disp_vel} show the mean metallicity (top) and the velocity dispersion (bottom) as a function of the projected elliptical distance in Sextans.
Here we use [B22~$+$~P22] only as most of the new members are located in much outer regions (black circles in Figure 2).
Each point represents the mean of one of the two quantities in a bin that is wide $1r_h$. In this plot, blue markers are the members from \citetalias{Battaglia22} and \citetalias{Pace22}, while black and olive circles represent data from \citet{Battaglia11} and \citet{Tolstoy25}, respectively. 
Another dataset would be \citet{Kirby10}, however it is not included in this work as their observed Sextans stars are mostly locate in the inner regions of the system. 

Recently, \citet{Tolstoy25} have re-analysed the Sextans stars from \citet{Battaglia11}, including  members from \citet{Walker23}. They discussed the presence of a double-peak profile in the Sextans' metallicity distribution, respectively at $\FeH\sim-3.0$ and $\FeH\sim-2.0$. These two populations have slightly different velocity dispersions, with the stars at $\FeH\sim-3.0$ being more kinematically hotter. Panel~E~of~Figure~
\ref{Fig:sextans} does not show trace of a double-peak metallicity profile, given the limited number of stars at the lowest-metallicities. However, we clearly see the radial gradients.

While the outskirts is more metal-poor and the metallicity gradient flattens out, compatible with null variations. The velocity dispersion increases with the distance  from Sextans' centre. Such behaviour is in agreement with the previous findings from \citet{Battaglia11} -- which has almost no members beyond $3r_h$--, and with those from \citet{Tolstoy25}. The outermost value of the velocity dispersion from \citet{Tolstoy25} is out of the y-axis range as it is scarcely populated, providing not useful values compared to our DESI value. 

\subsection{Velocity and metallicity profiles: comparison with cosmological zoom-in simulations}

The observed metallicity gradient and velocity dispersion profiles of Figure~\ref{Fig:disp_vel} are also in great agreement with our simulated dwarf galaxy (red lines -- a $+0.5$~dex has been added to the metallicity for a better visualisation, without modifying its trend).
The red lines represent the values derived from a Sextans-like dwarf satellite galaxy from our cosmological `zoom-in' hydrodynamical simulation that also includes detailed chemical evolution (Sestito et al., in prep.); the simulation code is based on Gadget-3 \citep{Springel05} including baryon physics described in \citet{Kobayashi07} and the latest nucleosynthesis yields from \citet{Kobayashi20b,Kobayashi20}. The initial condition is taken from \citet[][Aq-C-4]{Scannapieco12}, which forms a Milky Way-type galaxy; the detailed properties of this main galaxy has been analysed in \citet{Vincenzo20}.
The simulation also contains a few hundreds satellite galaxies, spanning a wide ranges of chemo-dynamical properties (Sestito et al. in prep.), from which we selected an initial pool of 8 systems with stellar masses from 4--$5\times10^5\rm{M_{\odot}}$, i.e. similar to Sextans' stellar mass. Half of the simulated dwarf galaxies experienced a merging event in their history, which have brought in approximately $\sim8$--13 percent of their stellar masses. The other half of the sample has no accreted stars. The simulated galaxy that best matches the observed metallicity and velocity dispersion trends of Sextans has not experienced a merging event nor is tidally perturbed by other nearby systems. \citet{Cicuendez18} suggested that Sextans might undergone an accretion event from the presence of a shell-like structure in the system north-east side, although the accreted mass was not estimated. The resolution of the Aq-C-4 suite does not permit to investigate accretion of systems with stellar masses $\lesssim10^4\rm{M_{\odot}}$. As regard tidal perturbations, Sextans has a distant Galactic pericentre ($\sim70$~kpc), also far from the Magellanic clouds \citep[$\gtrsim100$~kpc,][]{Battaglia22}, and thus it is unlikely to be gravitationally perturbed by these more massive galaxies. Similarly, the simulated galaxy is far from other systems ($\gtrsim100$~kpc) at the present day. The stellar mass of the simulated galaxy is also very similar to that of Sextans, i.e. $4.6\times 10^5$~M$_{\odot}$ (vs  $\sim\!4.4\times 10^5$~M$_{\odot}$ for Sextans, \citealt{Battaglia22}). On the star formation history, both  the simulated galaxy and Sextans had one single episode of star formation. The simulated galaxy has formed half of its stellar content within $\sim2.5$~Gyr, which is comparable to the value reported by \citet{Bettinelli18} for Sextans. 

As we can see from Figure~\ref{Fig:disp_vel}, the inner region of Sextans has a higher mean metallicity, where also the metallicity gradient is steeper ($\sim$\,$-0.25$~dex~r$_{h}^{-1}$ or $\sim$\,$-12\times 10^{-3}$~dex~arcmin$^{-1}$ or $\sim$\,$-0.36$~dex~kpc$^{-1}$)\footnote{Assuming a half-light radius of 695~pc \citep{Mcconnachie12} or of 21.4~arcmin \citep{Battaglia22}.}. In the outskirts ($r_h>3$), the plot shows a constant metallicity value around $\sim-2.5$. The gradient in metallicity found with DESI DR1 data in this work is comparable with the values derived using the data from \citet{Battaglia11} and \citet{Tolstoy25}, which are obtained from intermediate resolution spectroscopy of the VLT/FLAMES (R~$\sim6500$). Recently, \citet{Taibi22} discussed the metallicity gradient ranges in Local Group dwarf galaxies, using values from the literature. In \citet{Taibi22}, the reported value for Sextans' metallicity gradient is slightly steeper than those of other dwarf galaxies with similar luminosity, Galactic distance, stellar mass, gas mass, or star formation timescale. The value we obtain from DESI DR1 data is consistent with that reported by \citet{Taibi22} within $\sim\!1\sigma$, but is slightly less steep, which makes Sextans' value closer to the 95 percent confident interval obtained from all the Local Group DGs \citep[see Figures~3~and~4 of][]{Taibi22}.

In relation to the formation of radial metallicity gradients, one could assume that the oldest and most metal-poor stars form everywhere in the systems; the enriched gas cools and sinks in the inner regions while part of it is expelled from the outskirts due to the lower potential; eventually,  metal-rich stars form only in the inner regions \citep[e.g.][]{Zhang12, Hidalgo13,BenitezLlambay16,Revaz18}. This mechanism, called ``outside-in'' formation, is commonly observed among various Local Group dwarf galaxies \citep[e.g.][]{Battaglia11,Walker11,Vitali22,Tolstoy23,Sestito23scl,Sestito23Umi,Sestito24Sgr2,Arroyo-Polonio25,DESIdraco}. In addition, the recurrent star formation feedback might also push outwardly the oldest and most metal-poor stars from the centre. These will be investigated in our future work with our simulations.

In summary, Figures~\ref{Fig:sextans}~and~\ref{Fig:disp_vel} show a  metal-rich and kinematically colder population confined in the inner regions and a  hotter, dispersed, more metal-poor one in the outskirts. The best way to distinguish these scenarios would be to derive detailed elemental abundances of stars.

\subsection{Elemental abundances of the $\alpha-$elements}

\begin{figure*}
\includegraphics[width=1\textwidth]{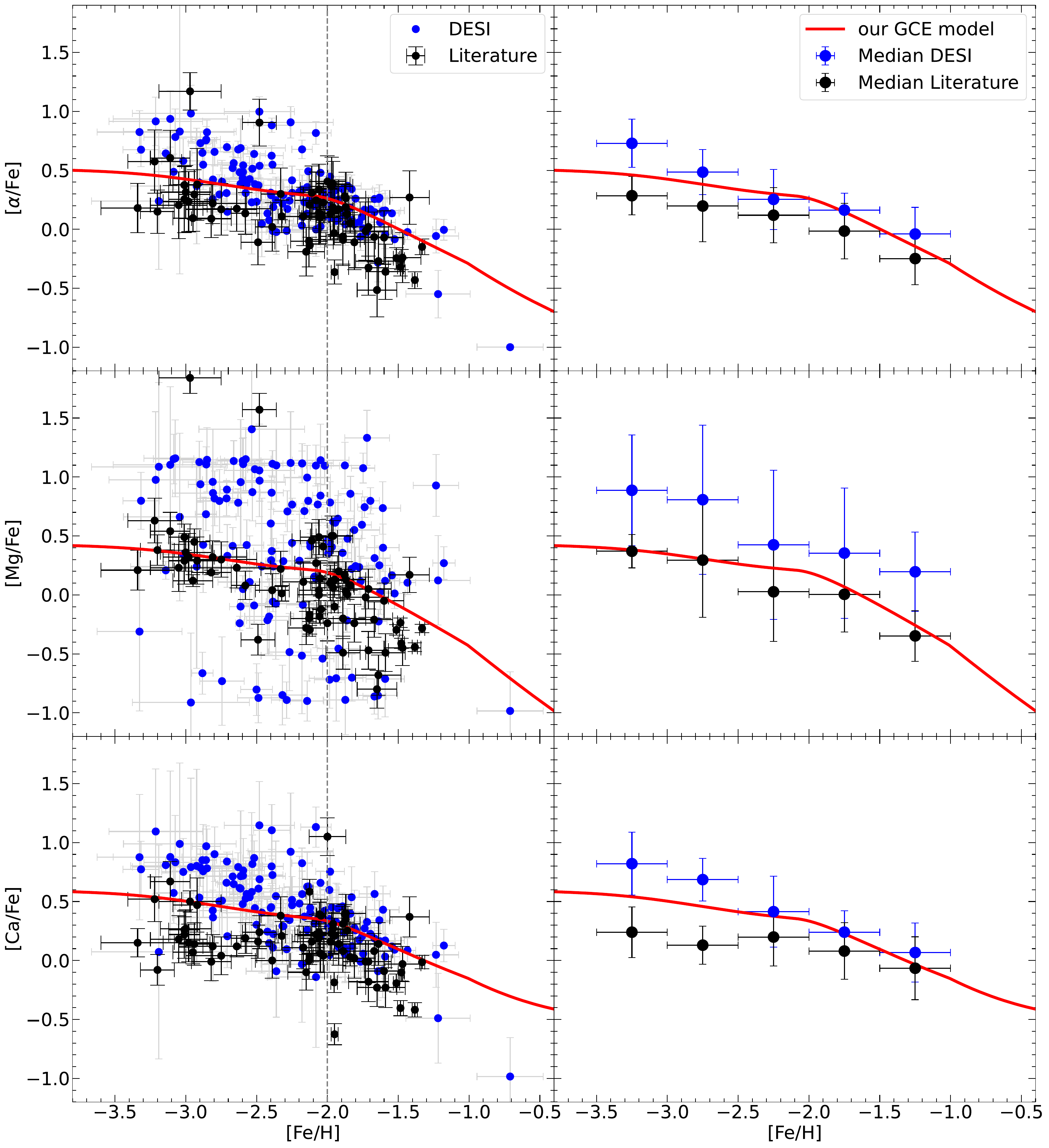}
\caption{Elemental abundances of $\alpha-$elements in Sextans for [$\alpha$/Fe] vs {\FeH} (top), [Mg/Fe] vs {\FeH} (middle), and [Ca/Fe] vs {\FeH} (bottom). DESI DR1  and their median values are shown with blue~$+$~grey errorbars (left panels) and blue circles (right panels), respectively. DESI stars with $\sigma_{\FeH} >0.5$~dex have been removed from the plot. Members stars are selected from \citetalias{Battaglia22} and \citetalias{Pace22}. For comparison, black circles (left panels) are high-resolution spectroscopic LTE elemental abundances taken from \citet{Mashonkina17b}; \citet{Lucchesi20} with their revised sample from \citet{Aoki09}; \citet{Theler20}; \citet{Roederer23}, and from APOGEE DR17 \citep{APOGEEDR17}, while their median values are shown with black circles (right panels). Our GCE model for Sextans  \citet{Kobayashi20b} is also shown with a red line, which is revised to match the position of the $\alpha-$knee as in \citet{Theler20}. For the literature and gce model values,  [$\alpha$/Fe] is defined as the mean between [Ca/Fe] and [Mg/Fe], while for DESI data this quantity is extracted by the pipeline. APOGEE DR17 data are selected imposing SNR~$>50$, and on-sky position, radial velocities and proper motion compatible with those of Sextans. Errorbars on the medians x-axis in the right panels denote the width of the bins. Vertical dashed grey lines mark the \FeH{} at which, below that, DESI [X/Fe] are supposed to be ``not reliable''.}
\label{Fig:chem_sex}
\end{figure*}

It has been shown that  DESI DR1 elemental abundances [X/Fe] might present a strong anticorrelation with metallicities \FeH, especially for stars with the lowest metallicities \citep{DESIstellarcat}. Nonetheless and with extreme caution, we show in Figure~\ref{Fig:chem_sex} the [$\alpha$/Fe] (top left panel), [Mg/Fe] (middle left panel), and  [Ca/Fe] (bottom left panel),  extracted by DESI DR1 pipeline (blue markers and grey errorbars), as a function of \FeH, which show a significant scatter. For comparison, high-resolution spectroscopic  values in local thermodynamic equilibrium (LTE, black circles) for 69 stars from \citet[][2 stars]{Mashonkina17b}, from \citet[][2 stars]{Lucchesi20}, from \citet[][6 stars]{Aoki09} re-analysed in \citet{Lucchesi20}, from \citet[][38 stars]{Theler20},   from \citet[][5 stars]{Roederer23}, and from APOGEE DR17 \citep[][16 stars]{APOGEEDR17} are also plotted in the left panels. In the right panels, median values of the elemental abundances for both DESI DR1 (blue) and the literature (black) are are shown, which do not agree very well.

Elemental abundances expected from our galactic chemical evolution (GCE) model (red line) are plotted in the right column of Figure~\ref{Fig:chem_sex}. 
The GCE code is described in \citet{Kobayashi00}, and the latest nucleosynthesis yields from \citet{Kobayashi20b,Kobayashi20} are included. 
For Type Ia supernovae (SNe Ia), both Chandrasekhar (Ch)-mass and sub-Ch mass explosions are included\footnote{Equation 2 of \citep{Kobayashi09} is used for both but with different secondary mass ranges; 0.835--1.9$M_\odot$ for single degenerate systems \citep{kobayashi15} and 1.8--7.95$M_\odot$ for double degenerate systems, depending on the metallicity.}.
The model parameters of star formation, inflow, outflows are basically chosen for Sextans in \citet{Kobayashi20b}, but are revised to match the position of the $\alpha-$knee, which should appear at $-2.0\lesssim\FeH\lesssim-1.5$ \citep{Theler20}. 
More specifically, we increased (i) the timescale for the star formation duration (from 100 to 200~Gyr), (ii)  the timescale for the outflow of gas (from 1.4 to 1.6~Gyr), and (iii) the contribution rate of sub-Ch~SNe~Ia by a factor of 1.5. The new version of the model is in excellent agreement with the high-resolution elemental abundances from the literature. To be noted, Mg has a steeper slope than Ca after the $\alpha-$knee ("the shin"), in both the GCE model and in the observed median [(Mg, Ca)/Fe] ratios. We interpret the different shin' slope as an evidence that sub-Ch~SNe~Ia contributed to the chemical evolution of Sextans, as this kind of SNe~Ia  produces more Ca than Mg \citep[e.g.][]{Kobayashi20b}.

For stars with $\FeH\gtrsim-2.0$ (on the right of the vertical dashed grey line), there seems to be an agreement in the  [$\alpha$/Fe] and [Ca/Fe] between DESI DR1 and the high-resolution estimates, also between their median values within $1\sigma$. [Mg/Fe] seems to be problematic as the distribution is more scattered in DESI. Literature data also show a trend, rather than a dispersion, in [($\alpha$, Mg, Ca)/Fe] at this metallicity regime. 
As also discussed in \citet{Theler20}, SNe Ia seem to start to contribute to the chemical enrichment of Sextans at these metallicities, which can explain the behaviour of the [X/Fe] from the literature. 
The slope at [Fe/H]~$\gtrsim-2.0$ is shallower for Ca than Mg, which is also expected in the SN~Ia nucleosynthesis \citep{Kobayashi20b}.
The position of the $\alpha-$knee at such \FeH{} is similar to those of other DGs \citep[$-2.5\lesssim\FeH\lesssim-1.5$, e.g.][]{Tolstoy09,Venn12,Hendricks14,Norris17,Reichert20,Theler20,Sestito23Umi,Vitali25}. In DGs, the onset of SNe~Ia happens at lower \FeH{} than in the Milky Way, as  the retention of metals in these relatively low-mass systems is less efficient than in the Galaxy \citep[e.g.][]{Venn04,Tolstoy09,Haywood2013,Kobayashi20b}. 

For $\FeH\lesssim-2.0$ (on the left of the vertical dashed grey line), DESI values   show an unrealistic scatter in all the three left panels (up to $\sim1.5$~dex for [Mg/Fe]). The extremely high or low values in [($\alpha$, Mg, Ca)/Fe] extracted by DESI DR1 in the very metal-poor regime are inconsistent with the GCE model and also the high-resolution literature data (except for 3 high-$\alpha$ stars, see more discussion in Appendix~\ref{AppSex}). In addition, the median values of [X/Fe] increases as metallicity decreases (right panels), presenting  a strong anticorrelation. The median values from DESI (blue circles) are systematically higher and with a steeper slope than the literature median values (black circles). The large scatter of [X/Fe] (especially for Mg), the  numerous stars with extreme [X/Fe], and the strong anticorrelation with metallicity would imply extreme caution for any physical interpretation of the DESI elemental abundances at the lowest metallicities.

\section{The most metal-poor stars}\label{sec:ump}

\begin{figure*}
\includegraphics[width=0.9\textwidth]{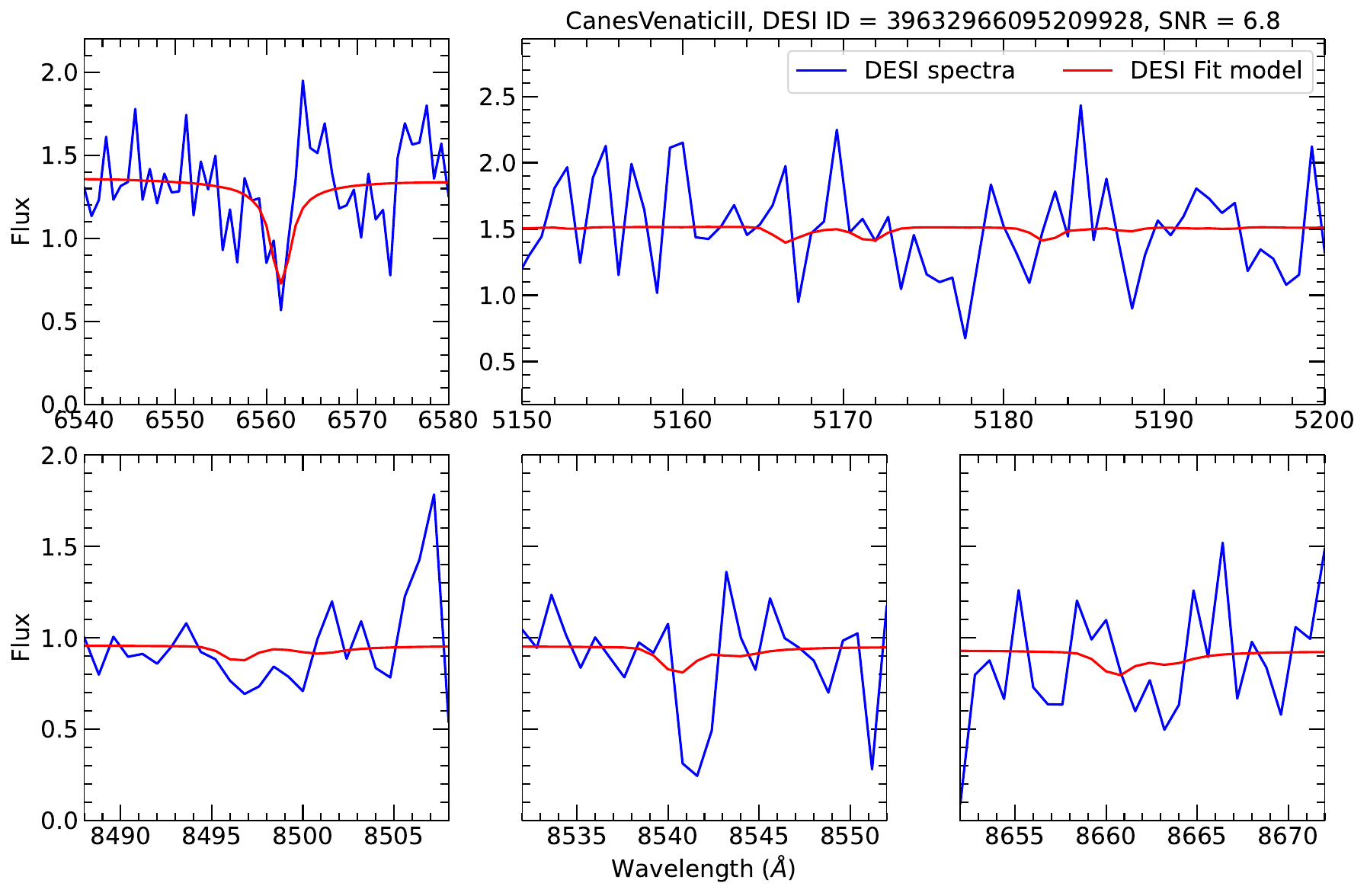}\\
\includegraphics[width=0.9\textwidth]{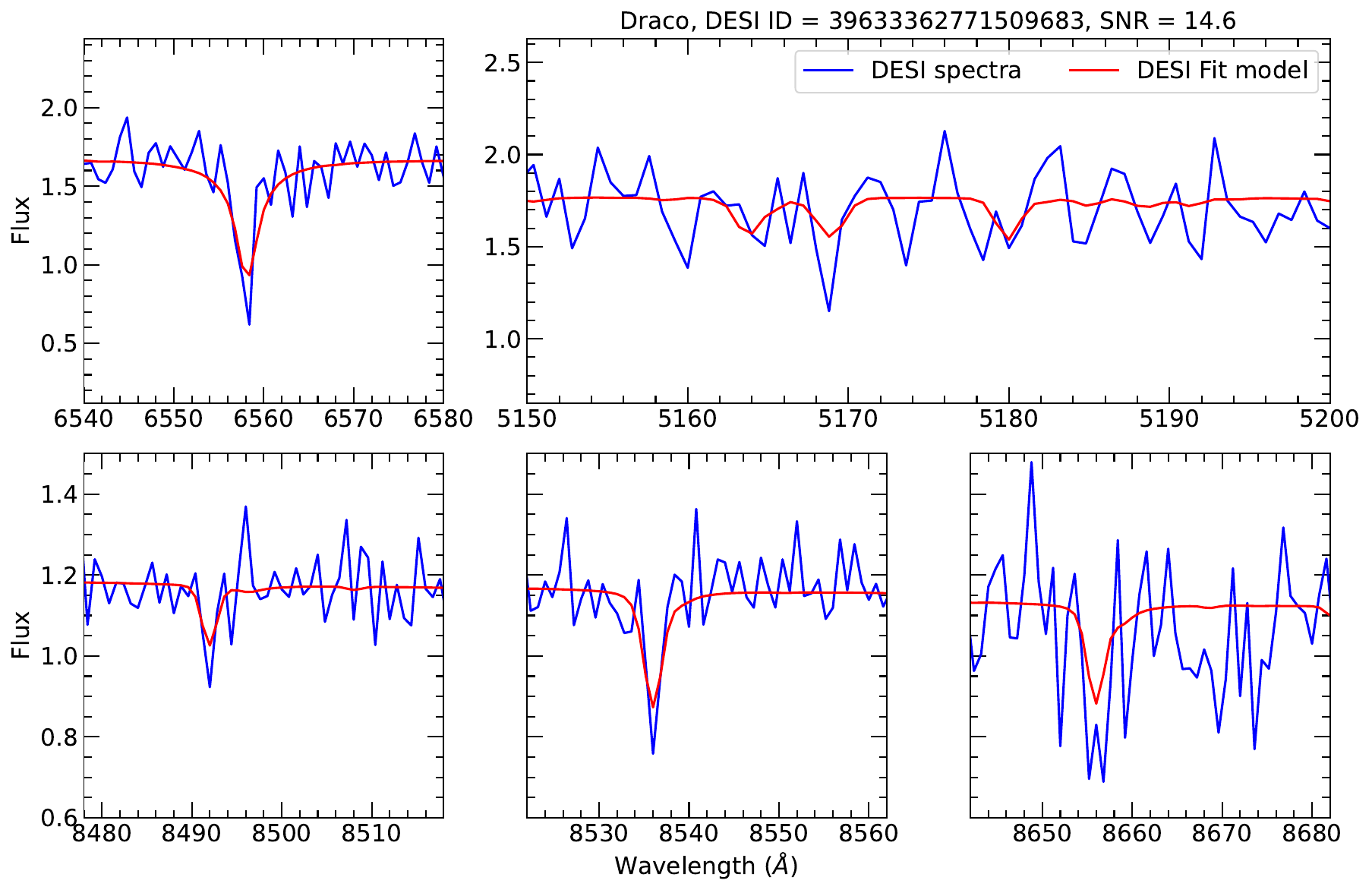}
\caption{Spectra of candidate UMP stars, one in Canes~Venatici~II and one in Draco. DESI spectra and their spectral fitting models are shown in blue and red lines, respectively. Spectra are not corrected for radial velocity shifting nor flux normalised. For each star, the H~$\alpha$ (top left), the Mg\ione{} Triplet (top right), and the Ca\ii{} Triplet (bottom row) spectral regions are displayed. Galaxy, DESI ID, and SNR are reported as title for each star.}
\label{Fig:ump}
\end{figure*}

\begin{table*}
    \centering 
    \begin{tabular}{lccccccc}
         Galaxy & DESI DR1 ID  & Gaia DR3 Source ID  & $\FeH_{\rm{cal}}$ & SNR &  \textsc{BESTGRID} & Catalogue   & Note \\  
         field & & & (dex) & @Ca\ii{} T&  & &  \\\hline
CVen~II & 39632966095209928 & 1515697326013099904 & $-4.30 \pm 0.63 $ &  6.8 & !$=$~\textsc{s\_rdesi1} & B22    & UMP candidate \\ \hline
Dra & 39633362771515905 & 1433204237051138304 & $-3.74 \pm 0.33 $  & 8.5 & !$=$~\textsc{s\_rdesi1} & B22   & UMP candidate  \\
Dra & 39633356299700349 & 1433055729967360128 & $-3.66 \pm 0.24 $ &  7.3 & !$=$~\textsc{s\_rdesi1} & B22/P22  & UMP candidate \\ 
Dra & 39633362771509683 & 1433203820439296640 & $-3.62 \pm 0.02$ & 14.6 &  $=$~\textsc{s\_rdesi1} & B22/P22   &  $\FeH_{\rm{CaT}} = -2.7\pm0.2$ \\\hline
Sex & 39627745013008688 & 3829135422249251968  & $-4.80 \pm 0.72 $  & 2.5 & !$=$~\textsc{s\_rdesi1} & B22 unc  & UMP candidate \\
Sex & 39627738977404585 & 3829118379819009024  & $-4.26 \pm 0.63 $  & 2.0 & !$=$~\textsc{s\_rdesi1} & B22 unc   & UMP candidate \\ 
Sex & 2305843037739159875 & 3829058177262521216 & $-3.61 \pm 0.10$ & 1.2 &  $=$~\textsc{s\_rdesi1} & B22/P22     & UMP candidate \\
Sex & 39627738981597619 & 3829065358447854592 & $-3.61 \pm 0.13$ & 2.4 &  $=$~\textsc{s\_rdesi1} & B22    & UMP candidate \\
Sex & 39627763140789618 & 3831439204052083584 & $-3.58 \pm 0.06$ & 7.3 &  $=$~\textsc{s\_rdesi1} & B22     & UMP candidate \\ \hline
\hline
IC1613 & 2851374429634560 & 2539072107188195584 & $ -4.99\pm 0.05$  & 1.6  & !$=$~\textsc{s\_rdesi1} & B22  & Hot star \\ \hline
M33 & 1152921504619431695 & 303363474380375936 & $-4.81 \pm 0.08 $ & 5.0 & !$=$~\textsc{s\_rdesi1} & B22    &  Hot star\\ \hline
CVen~I & 39632956167291660 & 1469959020983848704 & $-3.876 \pm 0.002$  &6.2 &  $=$~\textsc{s\_rdesi1} & New    & Galaxy \\ \hline
Dra & 39633359554480798 & 1433139292851749888  & $-4.03 \pm 0.08$ & 2.3 &  $=$~\textsc{s\_rdesi1} & B22/P22 unc    & Galaxy  \\
Dra & 39633356308088548& 1432939220389369984 & $-4.11 \pm 0.04$ & 4.0 &  $=$~\textsc{s\_rdesi1} & P22 unc    & Galaxy \\ \hline
Sex & 39627732933414594 & 3828973931478853632  & $-3.98 \pm 0.02$ & 3.7 &  $=$~\textsc{s\_rdesi1} & B22/P22  unc    & Galaxy \\
Sex & 39627732937606255 & 3829067007715281536  & $-3.67 \pm 0.11$ & 9.3 &  $=$~\textsc{s\_rdesi1} & B22/P22  unc    & Galaxy \\  \hline
    \end{tabular}
    \caption{Ultra metal-poor candidates in Local Group galaxies and contaminants. We report the DESI DR1 ID, the Gaia DR3 designation, the calibrated metallicity and its uncertainty, the derived signal-to-noise (SNR) ratio around the Ca\ii{} Triplet, the best grid flag adopted by DESI pipeline, the catalogue from which the object has been confirmed or unconfirmed (unc) as a member, and a column reporting whether the object is a UMP candidate or something else.}
    \label{tab:metalpoor}
\end{table*}

Ultra metal-poor (UMP, [Fe/H]~$<-4.0$) stars are thought to be likely polluted by one supernova event originated by the First stars \citep[e.g.][]{Frebel12,Nomoto13, Starkenburg18,Hartwig23,Bonifacio25}. 
 So far, less than 50 objects have been discovered in the Milky Way \citep{Sestito19,Limberg25}, and only three in dwarf galaxies, i.e. one in Sculptor \citep{Skuladottir21}, one in the Large Magellanic Cloud \citep{Chiti24}, and one in Pictor~II \citep{Chiti26}.
As UMP stars are extremely rare objects \citep[e.g.][]{Youakim17}, the discovery new UMP stars in the Milky Way and its satellites would be invaluable for studying the chemical enrichment in the early Universe. Low-resolution spectroscopic surveys such as DESI can provide good candidates.

However, as spectral lines in UMP stars are very weak \citep[e.g.][]{Starkenburg18,Lardo21,Skuladottir21,Chiti24}, automatic fitting pipelines of low-resolution spectra might not work properly on such objects, and thus we visually investigate the candidates as follows. 
In Figures~\ref{Fig:selexample}~and~\ref{Fig:sextans}, the most metal-poor stars are those possessing large uncertainties on the derived metallicities.
We gather all the stars (members and unconfirmed members of systems studied in this work) in all galaxies to assemble a sample of candidate UMP stars with $\FeH \lesssim -3.4$ (adding their uncertainty). We ease our selection criteria here from Section~\ref{sec:crit}, including stars with uncertainties on metallicity $\sigma_{\FeH}<1.0$ and stars with \textsc{BESTGRID}~$=$~\textsc{s\_rdesi1}, and removing the limit on the RV uncertainties. This allows us to find 16 objects, listed in Table~\ref{tab:metalpoor}. 
We then use the DESI retriever tool\footnote{\url{https://github.com/segasai/desi_retriever/}} to download the DESI spectra and the model fit of these 16 objects. Upon inspection of their spectra, 5 objects are found to be emission line galaxies, all of which have \textsc{BESTGRID}~$=$~\textsc{s\_rdesi1}. Two objects located in M33 and IC1613 are hot stars \citepalias[as discussed in][]{Battaglia22}, therefore they cannot be considered as old UMP stars.

The remaining 9 objects are stars and we inspect their spectra and derive the signal-to-noise ratio (SNR) around the Ca\ii{} Triplet. Examples of the DESI spectra (blue lines) and their spectral fitting models (red lines) for one star in Canes~Venatici~II and one in Draco are reported in Figure~\ref{Fig:ump}. The panels of Figure~\ref{Fig:ump} display the H~$\alpha$ (top left panel of each star), the Mg\ione{} Triplet (top right), the Ca\ii{} Triplet regions (bottom row).

The star in Canes~Venatici~II (top panels of Figure~\ref{Fig:ump}) has a barely detectable H~$\alpha$ line, while the Mg\ione{} Triplet and the Ca\ii{} Triplet lines are buried in the noise. 
Three UMP candidates are in Draco, two of them with \textsc{BESTGRID}~!$=$~\textsc{s\_rdesi1}. These two stars have H~$\alpha$ lines that are detectable and relatively well fitted by the models, while their Mg\ione{} Triplet and the Ca\ii{} Triplet lines are barely detectable if not washed away by the noise. The remaining star in Draco, which is shown in the bottom panels of Figure~\ref{Fig:ump}, has a SNR sufficient to detect H~$\alpha$ and Ca\ii{} Triplet lines.
There are 5 UMP candidates towards Sextans, two of them with \textsc{BESTGRID}~!$=$~\textsc{s\_rdesi1}. Both of them have very poor SNR, i.e. $\sim\!2.5$ and $\sim\!2.0$, which make their spectral fitting very unreliable. These two stars are Sextans candidate members from \citetalias{Battaglia22} and they did not pass our previous selection due to their large RV uncertainty of $\sim\!120$ and $\sim\!140$~km~s$^{-1}$, respectively. Their SNR is so low that only H~$\alpha$ is  detectable in one of them (DESI~ID~39627745013008688). The spectra of the remaining three stars also have low SNR values, although H~$\alpha$ and Ca\ii{} Triplet lines are detectable or barely detectable, while the Mg\ione{} Triplet is too weak compared to the noise.

Ca\ii{} Triplet lines have been widely used as a proxy for metallicity in giant stars \citep[e.g.][]{Armandroff91,Battaglia08,Starkenburg10,Carrera13,Longeard23} when dealing with low- or medium-resolution spectra. However, a SNR of at least $\sim\!15$ is needed to reliably estimate metallicity from the Ca\ii{} Triplet in the most metal-poor stars, as recently suggested by \citet{Tolstoy25}.
We find only one star in Draco that has a sufficient SNR close to this threshold.

\begin{figure*}
\includegraphics[width=1\textwidth]{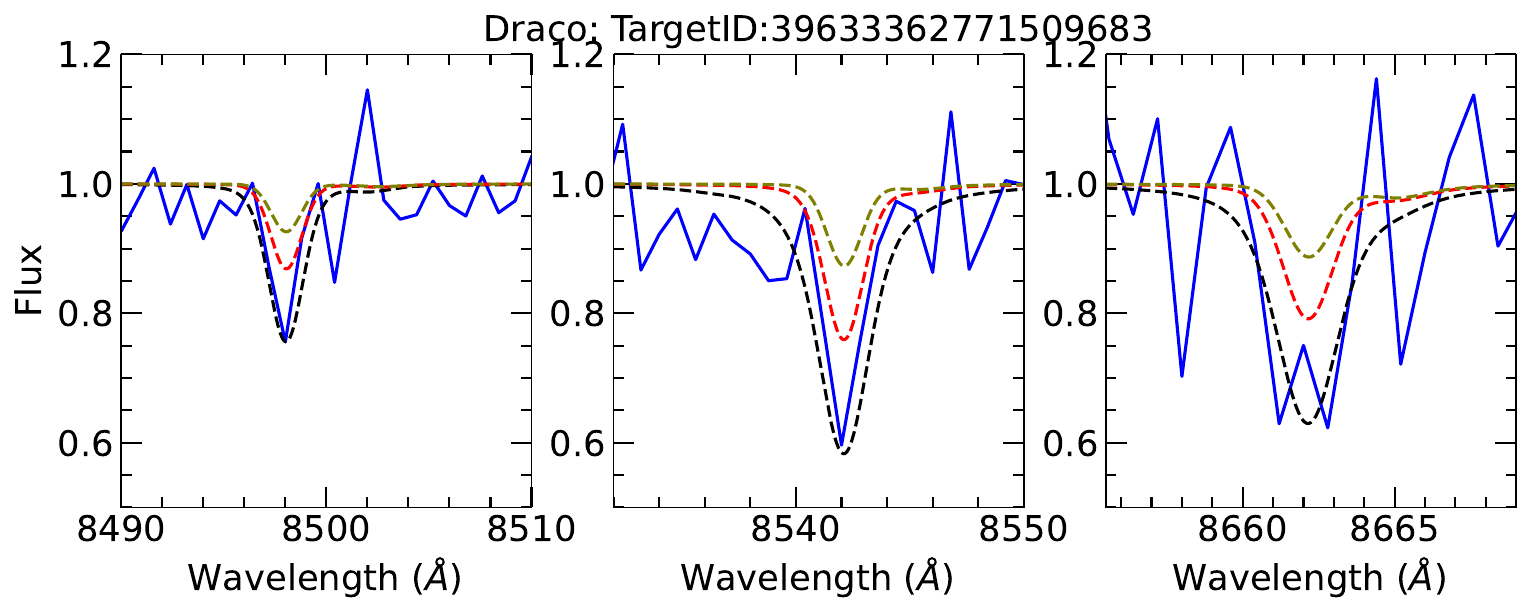}
\caption{Calcium~\ii{} Triplet lines for the UMP candidate in Draco with the highest SNR. Blue lines are the flux normalised DESI DR1 spectra. Olive, red, and black dashed lines are our synthetic spectra generated from MARCS models adopting [Ca/Fe]~$= +0.4$; and [Fe/H]~$=-4.0,-3.0$, and $-2.0$, respectively.}
\label{Fig:ump2}
\end{figure*}

Finally, we compare 
the DESI spectrum of this Draco UMP candidates with our synthetic spectra generated from MARCS  stellar atmospheres models \citep{Gustafsson08}, assuming [Ca/Fe]~$=+0.4$ \citep[in line with the expected trend for $\alpha$-elements, e.g.][]{Kobayashi20} and adopting the stellar parameters from DESI DR1. Figure~\ref{Fig:ump2} shows its normalised DESI spectrum 
(blue line), comparing to the synthetic spectra with $\FeH = -4.0, -3.0$, and $-2.0$ (olive, red, and black dashed lines, respectively). 
The first two components of the Ca\ii{} Triplet are narrower than the synthetic model with [Fe/H]~$\sim-2.0$, which might indicate a metallicity in the range $-3.0\lesssim\FeH\lesssim-2.0$. The third Ca\ii{} Triplet line is too noisy for a fair comparison. In addition, the equivalent widths (EW) of the first two Ca\ii{} Triplet components have been measured ($257$~and~$956$~m\AA, respectively) to then estimate the $\FeH$, adopting the polynomial relation from \citet{Carrera13}. Uncertainties on the EW ($20$~percent of relative uncertainty), on the Gaia photometry ($\pm0.05$~mag), on the distance of the star ($\pm10$~kpc), and on the polynomial coefficients from \citet{Carrera13} have been folded into a Monte Carlo randomisation to derive accurate uncertainties on the metallicity. The estimated metallicity  is $\FeH=-2.70\pm0.21$, which is within the metallicity range estimated using the synthetic spectra. The comparison with the synthetic spectra and the EW-based method suggest that the star is likely not an UMP, although it is a VMP star. Further higher SNR spectroscopic follow-up observations, even with low-resolution, are needed to further investigate the nature of the remaining 8 UMP candidates in Canes~Venatici~II, Draco, and Sextans that we find in DESI DR1.

\section{Conclusions}\label{sec:conclusions}
In this work, we explore the DESI DR1 catalogue to find stellar members in the Local Group dwarf galaxies. We cross-match DESI data with candidate members list from \citet{Battaglia22} and \citet{Pace22}, which are based on the exquisite Gaia proper motion, on sky position and photometry. Together with radial velocities from DESI, we find stellar members in  15 galaxies, including ultra-faint dwarfs, classical dwarf galaxies, and irregular dwarfs (see Table~\ref{tab:members}). We also show that the candidate selection algorithms based on Gaia alone are doing a great job in removing  foreground contaminations and in finding true stellar members  (see Figure~\ref{Fig:selexample} and Table~\ref{tab:members}), as the purity is $\gtrsim96$ percent. In addition, we provide new members, that had low probability membership in either \citetalias{Battaglia22} and \citetalias{Pace22}, but have DESI radial velocities in the range of five times the systemic velocity dispersion around the systemic RV values. These new members tend to be located in the outskirts of the systems. This indicates that Gaia-only based algorithms might miss the members in the extreme peripheries of dwarf galaxies, which can be recovered using spectroscopic information. In addition, two stars in the direction of DES~J0225$+$0304 have similar values of radial velocities and proper motions. If they are true members of DES~J0225$+$0304, then it leads to the first determination of the systemic radial velocity  (RV$_{\rm{sys}}=-150.0\pm7.0$~km~s$^{-1}$).

The systems with the highest number of members in DESI are Draco and Sextans, and as it was not published before, we discuss chemo-dynamical properties within Sextans in detail (Section~\ref{sec:gradients}). Thanks to DESI, the member stars reach the outskirts of Sextans, up to $\sim10\ r_h$ (Figure~\ref{Fig:sextans}). Our systemic velocity, systemic mean metallicity, and their dispersions are consistent with the literature's values \citep{Battaglia11,DESIstellarcat}. The most metal-poor stars have a slightly higher velocity dispersion than the  metal-rich group  (Figure~\ref{Fig:sextans}),  considering the [B22~$+$~P22] sample. Within the galaxy, the velocity dispersion is lower in the inner regions, a behaviour that is opposite for the mean metallicity of the system (Figure~\ref{Fig:disp_vel}). 
These radial gradients suggest that the  metal-rich population is kinematically colder and more confined in the inner parts of Sextans vs a kinematically hotter and more disperse metal-poor population. These results are also in line with the previous finding  \citep[e.g.][]{Battaglia11,Tolstoy25}, although they were not venturing in the extreme outskirts of Sextans.
We find that the metallicity gradient is steeper in the inner regions ($\sim-0.25$~dex~r$_{h}^{-1}$ or $\sim -12\times 10^{-3}$~dex~arcmin$^{-1}$ or $\sim -0.36$~dex~kpc$^{-1}$), while it flattens in the outskirts, reaching null values (see Figure~\ref{Fig:disp_vel}). The value of the metallicity gradient from DESI data and its behaviour are in agreement with previous findings using medium resolution spectroscopic analysis \citep{Battaglia11,Tolstoy25}.
Moreover, the observational results are in good agreement with our chemo-dynamical simulation of a dwarf satellite galaxy that had no merger or tidal perturbation.

DESI DR1 also provides elemental abundances for some $\alpha-$elements (see Figure~\ref{Fig:chem_sex}). The [($\alpha$, Mg, Ca)/Fe] for Sextans stars with $\FeH\gtrsim-2.0$ are in line with the high-resolution spectroscopic literature data  \citep{Aoki09,Mashonkina17b,Lucchesi20,Theler20,Roederer23}, however Mg display a large scatter in its elemental abundance. On the other hand, lower-metallicity stars show a large scatter of [($\alpha$, Mg, Ca)/Fe] (up to $\sim2.5$~dex for Mg) and an anticorrelation with \FeH, which is not seen in literature data, nor predicted by GCE models \citep{Kobayashi20b}. Our revised GCE model for Sextans is in excellent agreement with the median [X/Fe] values of the literature data, reproducing the $\alpha-$knee at [Fe/H]~$\sim-2.0$, consistent with previous works \citep[e.g.,][]{Theler20}.
As warned by \citet{DESIstellarcat}, the DESI's elemental abundance measurements (e.g. Mg and Ca) need updating; these elemental abundances are extremely useful for understanding the origin of each satellite galaxies.

Finally, we look for UMP candidate stars in the direction of the 15 Local Group galaxies observed by DESI DR1, selecting those stars  with $\FeH\lesssim-3.5$ and $\sigma_{\FeH}\lesssim1.0$~dex (Table~\ref{tab:metalpoor}).
After careful inspection of their spectra (see Figure~\ref{Fig:ump}~and~\ref{Fig:ump2}), we find potentially 8 UMP candidates in Canes~Venatici~II (1 stars), Draco (2 stars) and Sextans (5 stars). Given their low SNR  spectra, we are not able to firmly confirm or deny their UMP nature. High SNR spectroscopic observations (SNR~$\gtrsim15$), even with low-resolution, are needed for confirmation; new UMP members will be extremely valuable for understanding the chemical enrichment in the early Universe.

We are living at the beginning of an era in which various MOS instruments and surveys will observe thousands of stars in a glance for millions of objects in the Local Group. Subaru's PFS \citep{PFS,PFSGA} and DESI already started acquiring data, while WEAVE \citep{WEAVE24} and 4MOST \citep{deJong19} will soon start to operate. Similarly, in the next couple of decades, HRMOS \citep{HRMOS} and WST \citep{WST} will also play a crucial role in Galactic archaeology for measuring more accurate elemental abundances. 
To compare with theoretical predictions, it is necessary to put on the same scale the elemental abundances derived from the pipelines of the various MOS surveys, which can be done using those systems or stars observed by multiple surveys. For instance Sextans, studied here in details, or Draco, Ursa Minor are being observed by PFS and will be in the WEAVE footprint, as well as already present in DESI DR1. 
Elemental abundances in the most metal-poor stars are challenging to be derived, and dedicated pipelines within MOS surveys should be employed,
as shown in this work.

\section*{Acknowledgements}
The authors thank the anonymous referee for their insightful comments, which helped in improving the quality of the draft.
FS and CK acknowledge funding from the UK Science and Technology Facilities Council through grant ST/Y001443/1. 
This work used the DiRAC Memory Intensive service (Cosma8 / Cosma7 / Cosma6 [*]) at Durham University, managed by the Institute for Computational Cosmology on behalf of the STFC DiRAC HPC Facility (www.dirac.ac.uk). The DiRAC service at Durham was funded by BEIS, UKRI and STFC capital funding, Durham University and STFC operations grants. DiRAC is part of the UKRI Digital Research Infrastructure.

This research used data obtained with the Dark Energy Spectroscopic Instrument (DESI). DESI construction and operations is managed by the Lawrence Berkeley National Laboratory. This material is based upon work supported by the U.S. Department of Energy, Office of Science, Office of High-Energy Physics, under Contract No. DE–AC02–05CH11231, and by the National Energy Research Scientific Computing Center, a DOE Office of Science User Facility under the same contract. Additional support for DESI was provided by the U.S. National Science Foundation (NSF), Division of Astronomical Sciences under Contract No. AST-0950945 to the NSF’s National Optical-Infrared Astronomy Research Laboratory; the Science and Technology Facilities Council of the United Kingdom; the Gordon and Betty Moore Foundation; the Heising-Simons Foundation; the French Alternative Energies and Atomic Energy Commission (CEA); the National Council of Humanities, Science and Technology of Mexico (CONAHCYT); the Ministry of Science and Innovation of Spain (MICINN), and by the DESI Member Institutions: www.desi.lbl.gov/collaborating-institutions. The DESI collaboration is honored to be permitted to conduct scientific research on I’oligam Du’ag (Kitt Peak), a mountain with particular significance to the Tohono O’odham Nation. Any opinions, findings, and conclusions or recommendations expressed in this material are those of the author(s) and do not necessarily reflect the views of the U.S. National Science Foundation, the U.S. Department of Energy, or any of the listed funding agencies.

This work has made use of data from the European Space Agency (ESA) mission {\it Gaia} (\url{https://www.cosmos.esa.int/gaia}), processed by the {\it Gaia} Data Processing and Analysis Consortium (DPAC, \url{https://www.cosmos.esa.int/web/gaia/dpac/consortium}). Funding for the DPAC has been provided by national institutions, in particular the institutions participating in the {\it Gaia} Multilateral Agreement.

This research has made use of the SIMBAD database, operated
at CDS, Strasbourg, France \citep{Wenger00}. This work made
extensive use of TOPCAT \citep{Taylor05}.

\section*{Data Availability}
DESI DR1 data are publicly available, as well as candidate members from \citetalias{Battaglia22} and \citetalias{Pace22}. The final list of members, which their number is reported in Table~\ref{tab:members}, will be publicly available as supplementary online material and at the CDS (\url{cdsarc.cds.unistra.fr}), once the paper is published. The original galactic chemical evolution model from \citet{Kobayashi20b} is available at \url{https://star.herts.ac.uk/~chiaki/gce}, while the revised version is available upon reasonable request.



\bibliographystyle{mnras}
\bibliography{desi} 



\appendix

\section{High-$\alpha$ stars in Sextans}\label{AppSex}
Figure~\ref{Fig:chem_sex} shows the presence of DESI DR1 stars with unusual high abundances of $\alpha-$elements. 
Stars with extremely high-$\alpha$ measurements are very rare among the observed stars. Given the anti-correlation of the DESI elemental abundances with metallicity, we retain unlikely the true high-$\alpha$ nature of these stars.

In the Sextans' literature, there are two stars from \citet{Theler20} and one from \citet{Roederer23} that stand out from the Sextans chemical distribution. One star, S05$-$95 \citep{Theler20}, is extremely Mg-rich, with [Mg/Fe]~$\sim+1.57$, although normal in Ca for that metallicity ([Ca/Fe]~$\sim+0.25$). The authors reported a low SNR ($\lesssim10$) for the spectrum of this star, also warning that [Mg/Fe] is measured only from one Mg~I line \citep[5528.410 \AA{},][]{Theler20}. Another Mg-rich star is present in Sextans, J1008$+$0001 \citep{Roederer23}, with [Mg/Fe]~$\sim+1.84$ and [Ca/Fe]~$\sim+0.5$. Extremely Mg-enriched stars do exist in Local Group dwarf galaxies, however, they are  rare, e.g. 2 out of 61 observed stars in Sextans and 2 out of 45 observed stars in Ursa Minor \citep{Sestito23Umi}.

The findings of such Mg-enhanced might be more frequent in low-mass ultra-faint dwarf galaxies, as also shown in \citet[][e.g. BootesI, Coma Berenices, ReticulumII, SegueI, Ursa MajorII]{Ji19}. This offers an explanation: the Mg-enhancement might be caused by inhomogeneous mixing star formation \citep[e.g.][]{Frebel12}. Alternatively, in case carbon is also enhanced, there are two scenarios. One related to the enrichment from a companion due to mass transfer \citep[e.g.][]{Stancliffe09}, which would also increase s-process elemental abundances, observed as so-called CEMP-s stars \citep{Beers05}; another scenario is associated with enrichment from the primordial core-collapse supernovae \citep{Norris13}, observed as so-called CEMP-no stars if neutron-capture elements are not enhanced \citep{Beers05}. No firm conclusions can be reached for S05$-$95, as elemental abundances for carbon are lacking. J1008$+$0001 is  strongly enriched in C, Na, Si, and K, while it is deficient in the neutron-capture elements Sr, Ba and Eu. \citet{Roederer23} classify this star as the first and, currently, the only  CEMP-no discovered in Sextans.  

The third star, S08$-$113  \citep{Theler20}, is Ca-rich with [Ca/Fe]~$\sim+1.05$ (from 12 Ca~I lines), while it is Mg-poor, with [Mg/Fe]~$\sim-0.24$. Similarly, the spectrum of this star had a low SNR, therefore, these elemental abundances should be taken with `the grain of salt'. This combination of abundances might imply a high [Ca/Mg] ratio. One might think that such a high [Ca/Mg] ratio might be explained by the enrichment from pair-instability supernovae (PISNe; e.g. \citealt{Nomoto13}), however, to firmly assess the nature of this star's progenitor, elemental abundances of the key elements of PISNe (Al, Na, Si, Zn, Cu etc.) must be derived \citep[e.g.][]{Heger02,Umeda02,Nomoto13,Takahashi18,Umeda24}.


\bsp	
\label{lastpage}
\end{document}